\documentclass[apjl, iop]{emulateapj}

\slugcomment{Submitted to ApJ}

\usepackage{color}

\shorttitle{Observations of SN IIP 2013am}
\shortauthors{Jujia Zhang et al.}

\begin{document}

\title{Optical and Ultraviolet Observations of A Low-Velocity Type II-Plateau Supernova 2013am in M65}

\author{Jujia Zhang\altaffilmark{1,2,3}, Xiaofeng Wang\altaffilmark{4},
Paolo A. Mazzali\altaffilmark{5}, Jinming Bai\altaffilmark{1,3},
Tianmeng Zhang\altaffilmark{6,7}, David Bersier\altaffilmark{5},
Fang Huang\altaffilmark{4,8}, Yufeng Fan\altaffilmark{1,2,3}, Jun Mo\altaffilmark{4},
Jianguo Wang\altaffilmark{1,3,9}, Weimin Yi\altaffilmark{1,2,3},
Chuanjun Wang\altaffilmark{1,2,3}, Yuxin Xin\altaffilmark{1,3},
Liangchang\altaffilmark{1,3}, Xiliang Zhang\altaffilmark{1,3},
Baoli Lun\altaffilmark{1,3},  Xueli Wang\altaffilmark{1,3},
Shousheng He\altaffilmark{1,3}, and Emma S. Walker\altaffilmark{10}}
\altaffiltext{1}{Yunnan Observatories (YNAO), Chinese Academy of Sciences, Kunming 650216, China. (jujia@ynao.ac.cn)}
\altaffiltext{2}{University of Chinese Academy of Sciences, Chinese Academy of Sciences, Beijing 100049, China.}
\altaffiltext{3}{Key Laboratory for the Structure and Evolution of Celestial Objects, Chinese Academy of Sciences, Kunming 650216, China. (baijinming@ynao.ac.cn)}
\altaffiltext{4}{Physics Department and Tsinghua Center for Astrophysics (THCA), Tsinghua University, Beijing 100084, China. (wang\_xf@mail.tsinghua.edu.cn)}
\altaffiltext{5}{Astrophysics Research Institute, Liverpool John Moores University, Liverpool Science Park, 146 Brownlow Hill, Liverpool L3 5RF, UK}
\altaffiltext{6}{National Astronomical Observatories of China (NAOC), Chinese Academy of Sciences, Beijing 100012, China.}
\altaffiltext{7}{Key Laboratory of Optical Astronomy, National Astronomical Observatories, Chinese Academy of Sciences, Beijing 100012, China.}
\altaffiltext{8}{Astronomy Department, Beijing Normal University, Beijing 100875, China}
\altaffiltext{9}{Key Laboratory for Research in Galaxies and Cosmology, Chinese Academy of Science, 96 JinZhai Road, Hefei 230026, Anhui, China}
\altaffiltext{10}{Department of Physics, Yale University, New Haven, CT 06520-8121, USA}
\begin{abstract}
Optical and ultraviolet observations for the nearby type II-plateau supernova (SN IIP) 2013am in the nearby spiral galaxy M65 are presented in this paper. The early spectra are characterized by relatively narrow P-Cygni features, with ejecta velocities much lower than observed in normal SNe IIP (i.e., $\sim$2000 km s$^{-1}$ vs. $\sim$5000 km $^{-1}$ in the middle of the plateau phase). Moreover, prominent \ion{Ca}{2} absorptions are also detected in SN 2013am at relatively early phases. These spectral features are reminiscent of those seen in the low-velocity and low-luminosity SN IIP 2005cs. However, SN 2013am exhibits different photometric properties, having shorter plateau phases and brighter light-curve tails if compared to SN 2005cs. Adopting $R_{V}$=3.1 and a mean value of total reddening derived from the photometric and spectroscopic methods(i.e., $E(B-V)=0.55\pm$0.19 mag),  we find that SN 2013am may have reached an absolute $V$-band peak magnitude of $-15.83\pm0.71$ mag, and produced a $^{56}$Ni mass of $0.016^{+0.010}_{-0.006}$M$_{\sun}$ in the explosion. These parameters are close to those derived for SN 2008in and SN 2009N which have been regarded as ``gap-filler" objects linking the faint SNe IIP to the normal ones. This indicates that some low-velocity SNe IIP may not necessarily result from the low-energetic explosions, and the low expansion velocities could be due to a lower metallicity of the progenitor stars, a larger envelope mass ejected in the explosion, or that is was observed at an angle that is away from the polar direction.

\end{abstract}

\keywords{supernovae:general - supernovae: Individual (SN 2013am)}

\section{Introduction}
\label{sect:Intro}

Type II-plateau Supernovae (SNe IIP) represent a major group of core-collapse supernovae (CC SNe, see \citealp{CC09} for a review), characterized by a nearly constant optical luminosity for about 3-4 months after the peak. Special attentions have been paid to the SNe IIP showing low ejecta velocities (hereafter LV SNe IIP) because of their unusual properties in some aspects. The well-studied example include SN 1997D \citep{Tur97D, Be97D} and SN 2005cs \citep{05csa, 05csb}. These LV SNe IIP generally show a low-luminosity plateau lasting for about 100 days, followed by an under-luminous exponential tail at late times. Their spectra show prominent but narrower P-Cygni profiles. \citet{Past04} suggested that these LV objects are at the faint end of the luminosity distribution of SNe IIP.

\citet{Spiro14} recently presented observations of five faint SNe IIP (SNe 1999gn, 2002gd, 2003Z, 2004eg, and 2006ov). Combining with the published SNe IIP with low luminosities (e.g., SN 1994N, SN 1997D, SN 1999br, SN 1999eu, SN 2001dc, \citealp{Tur97D},  \citealp{Be97D},  \citealp{Past04}; SN 2005cs, \citealp{05csa,05csb}; SN 2008in, \citealp{Roy11}), they found that all faint SNe IIP have ejecta velocities below $\sim$1000 km s$^{-1}$ at the end of the plateau, and the $^{56}$Ni mass ejected in the explosion is smaller than 0.01 M$_{\sun}$. It is also suggested that those subluminous SNe IIP with low ejecta velocities probably originate from less energetic explosions of intermediate-mass stars, with an initial mass in the range of 10-15M$_{\sun}$. On the other hand, the `gap-filler' events like SN 2008in \citep{Roy11} and SN 2009N \citep{2009N} were recently reported, which showed a striking resemblance to the faint event SN 1997D but produced more $^{56}$Ni, i.e., $\sim$0.015 M$_{\sun}$ for SN 2008in and $\sim$0.02 M$_{\sun}$ SN 2009N.

In this paper we present optical and ultraviolet (UV) data of another low-velocity object SN 2013am, obtained through an observational campaign that lasted for about one year on several telescopes. The spectral evolution of SN 2013am shows an overall similarity to that of SN 2005cs which is known for its low expansion velocity and low luminosity. Observations and data reductions are described in Section \ref{sect:obs}, the UV and optical light and color curves are presented in Section \ref{sect:LC}, while the spectral evolution is shown in Section \ref{sect:Spe}. In Section 5, we estimate the explosion parameters such as the absolute magnitudes, quasi-bolometric light curve, and the $^{56}$Ni mass ejected in the explosion. A brief discussion about diversities of SNe IIP is also given in Section \ref{sect:Discu}. Finally, a summary is given in Section \ref{sect:con}.

\begin{figure}[!th]
\centering
\includegraphics[width=8cm,angle=0]{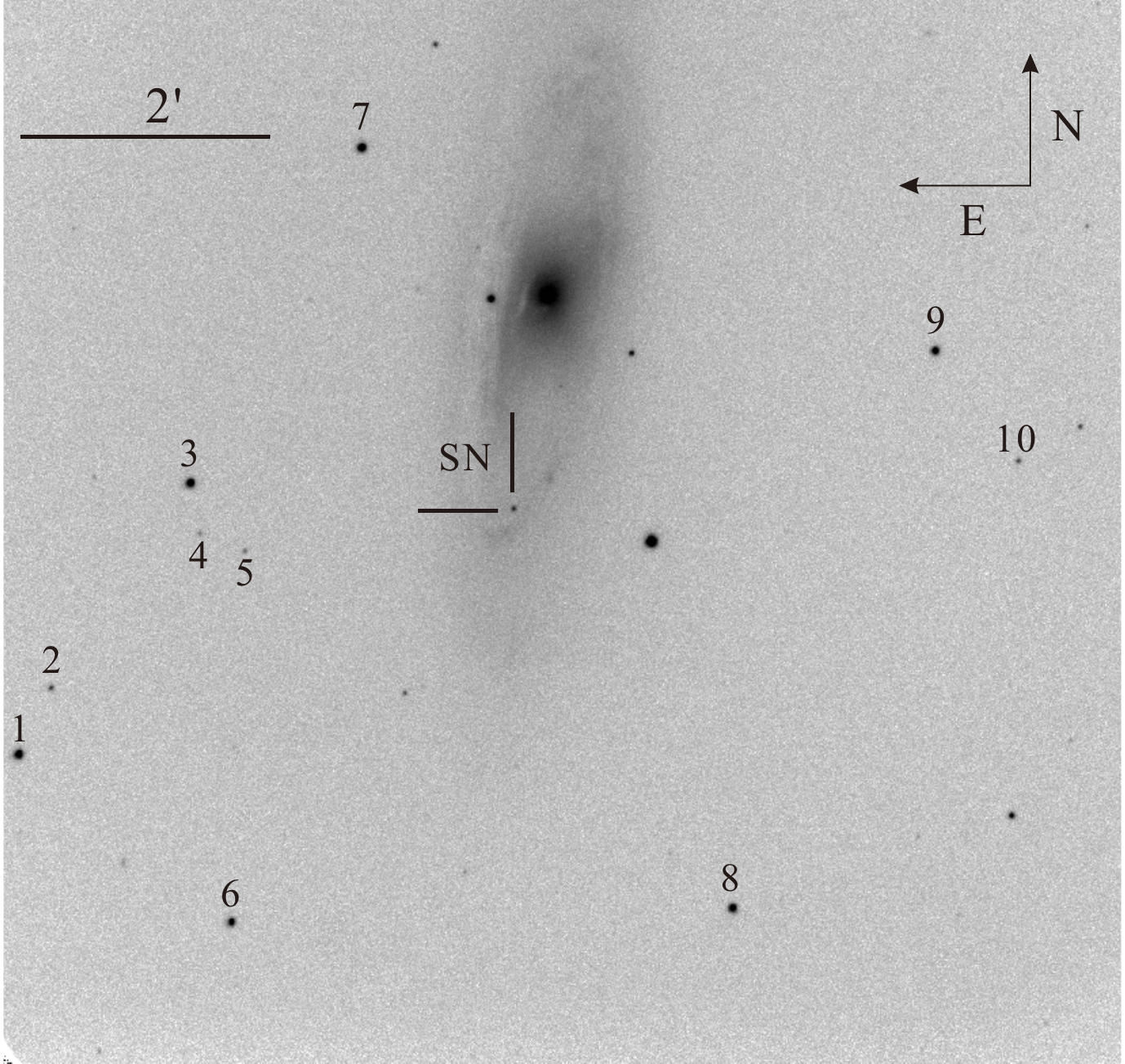}
 \caption{SN 2013am in M65. The $R$-band image was taken on Mar. 23 2013 by the 2.4-m telescope at Lijiang Observatory.
 The supernova and ten local reference stars are marked. North is up and East is left.}
\label{<img>}
\end{figure}

\section{Observations and Data Reduction}
\label{sect:obs}

SN 2013am was discovered at roughly a magnitude of 15.6 mag on Mar. 21.64 UT 2013 (UT is used throughout this paper) in an unfiltered image of the spiral galaxy M65 (NGC 3623) taken using a 0.35-m f/11.0 Schmidt-Cassegrain telescope \citep{Suga13}. Its coordinates are RA = 11h18m56s.95, Dec = $+13^\circ$03\arcmin49.4\arcsec (J2000). The SN is located 15\arcsec~east and 102\arcsec~south of the center of the host (see Figure \ref{<img>}).
This object was classified as a young type IIP based on a spectrum obtained on Mar.\ 22.84 2013 \citep{Be13id}.  The SN was also independently discovered by the Palomar Transient Factory (PTF) on Mar. 23.94 2013 (iPTF13aaz, \citealp{ATel4910}).  Nothing was detected at this location in the PTF $r$-band image taken on Mar. 19.34 2013, with an upper limit of 21.36 mag. This allows us to estimate the explosion time to be around Mar. 20.5 2013 (JD = 2456372.0$\pm$1.0). This epoch is thus assumed as the explosion date in this paper. Our observations started on Mar. 22.85 UT 2013,  $\sim$ +2.4 days after explosion.

\subsection{Photometry}
\subsubsection{Ground-based Observation}
\label{subsect:Ground}

The $UBVRI$ photometry of SN 2013am was obtained with the 2.4-m telescope at Li-Jiang observatory (hereinafter LJT, see \citealp{FYF} for a detailed
description), Yunnan observatories (YNAO), and the Yunnan Faint Object Spectrograph and Camera (YFOSC, see \citealp{zhang12,JJzhang14} for detailed descriptions). Optical photometry was also obtained with the Liverpool 2.0-m telescope (hereinafter LT), which is located at La Palma and is owned and operated by the Astrophysics Research Institute of Liverpool John Moores University.  Data were also collected with the Tsinghua-NAOC 80-cm telescope (hereinafter TNT, see Wang et al. 2008 and Huang et al. 2012 for detailed descriptions) at Xing-Long Station of National Astronomical Observatories (NAOC), China.

All CCD images were reduced using the IRAF\footnote{IRAF, the Image Reduction and Analysis Facility, is distributed by the National Optical Astronomy
Observatory, which is operated by the Association of Universities for Research in Astronomy(AURA), Inc. under cooperative agreement with the National Science Foundation(NSF).} standard procedure, including corrections for bias, flat field, and removal of cosmic rays.  The standard point-spread-function (PSF) fitting method from the IRAF DAOPHOT package \citep{Stetson} was used to measure the instrumental magnitudes for both the SN and local standard stars. These $UBVRI$ magnitudes were then converted to the standard Johnson $UBV$ \citep{Johnson} and Cousins $RI$ \citep{Cousins} systems through transformations established by observing \citet{Landolt} standards during photometric nights. The average value of the photometric zeropoints, determined on three photometric nights, was used to calibrate the local standard stars in the field of SN 2013am. Table \ref{Tab:Photo_stand} lists the standard $UBVRI$ magnitudes and the corresponding uncertainties of 10 local standard stars as labeled in Figure \ref{<img>}. The magnitudes of these stars were then used to transform the instrumental magnitudes of SN 2013am to those of the standard $UBVRI$ system. The $g$-, $r$-, and $i$-band magnitudes of the standard  stars listed in Table \ref{Tab:Photo_stand} are from  the Sloan Digital Sky Survey data release 10
(SDSS--DR10)\footnote{http://skyserver.sdss3.org/public/en/tools/chart/navi.aspx}. These $gri$-band magnitudes were converted into $BVRI$ magnitudes in light of the relations given by \citet{ugrizT} and \citet{Jordi06}. Note that those transformation relations were derived from normal stars, which may introduce systematic errors when applying to the photometry of supernovae. The final results of the $UBVRI$ magnitudes of SN 2013am are listed in Table \ref{Tab:GDPho}, and the $gri$-band magnitudes are listed in Table \ref{Tab:gri}.

\begin{deluxetable*}{ccccccccc}[!th]
\tablecaption{Magnitudes of the Photometric Standards in the Field of SN 2013am}
\tablewidth{0pt}
\tablehead{\colhead{Star} &  \colhead{$U$(mag)} & \colhead{$B$(mag)} & \colhead{$V$(mag)} & \colhead{$R$(mag)} & \colhead{$I$(mag)} & \colhead{$g$(mag)\tablenotemark{a}} & \colhead{$r$(mag)\tablenotemark{a}} & \colhead{$i$(mag)\tablenotemark{a}} }
\startdata
1	&	16.75(04)	&	15.42(02)	&	14.08(03)	&	13.28(03)	&	12.59(03)&	14.83(01)	&	13.66(01)	&	14.51(01)	\\
2	&	19.49(04)	&	18.74(03)	&	17.10(02)	&	16.08(03)	&	14.88(02)&	18.03(01)	&	16.56(01)	&	15.66(01)	\\
3	&	14.64(03)	&	14.47(03)	&	13.78(03)	&	13.39(04)	&	13.04(02)&	14.07(01)	&	13.64(01)	&	13.51(01)	\\
4	&	18.84(04)	&	18.46(02)	&	17.70(03)	&	17.25(04)	&	16.78(04)&	18.05(01)	&	17.49(01)	&	17.27(01)	\\
5	&	18.33(05)	&	18.27(04)	&	17.52(03)	&	17.08(05)	&	16.68(04)&	17.84(01)	&	17.38(01)	&	17.19(01)	\\
6	&	15.58(03)	&	15.35(02)	&	14.54(04)	&	14.07(05)	&	13.60(03)&	14.93(01)	&	14.36(01)	&	14.14(01)	\\
7	&	14.06(04)	&	14.17(04)	&	13.58(03)	&	13.23(02)	&	12.90(02)&	13.81(01)	&	13.48(01)	&	13.39(01)	\\
8	&	17.17(05)	&	16.02(02)	&	14.58(02)	&	13.74(03)	&	12.94(02)&	15.33(01)	&	14.04(01)	&	13.52(01)	\\
9	&	14.67(05)	&	14.71(03)	&	14.14(02)	&	13.82(04)	&	13.54(03)&	14.37(01)	&	14.05(01)	&	13.97(01)	\\
10	&	18.70(03)	&	18.02(03)	&	16.87(02)	&	16.30(03)	&	15.72(02)&	17.44(01)	&	16.56(01)	&	16.24(01)
\enddata
\tablecomments{ Uncertainties (numbers in brackets), in units of 0.01 mag, are 1 $\sigma$. See Figure~\ref{<img>} for the finder chart of SN 2013am and the comparison stars.}
\tablenotetext{a}{Measurements from SDSS--DR10.}
\label{Tab:Photo_stand}

\end{deluxetable*}

\subsubsection{Space-based Observation}
\label{subsect:Space}

SN 2013am was also observed with the UVOT \citep{UVOT05} on board the $Swift$ satellite \citep{Swift04}, spanning from $t \approx +3$ days to $t \approx +30$ days from the explosion. The photometric observations were performed in three UV filters ($uvw2$, $uvm2$, and $uvw1$) and three broadband optical filters ($U$, $B$, and $V$). These $Swift$ images were reduced using the ``uvotsource" program of the HEASoft (the High Energy Astrophysics Software\footnote{http://www.swift.ac.uk/analysis/software.php}) with the latest $Swift$ Calibration Database\footnote{http://heasarc.gsfc.nasa.gov/docs/heasarc/caldb/swift/}. Aperture photometry was adopted in the imaging reduction as the PSF profile is slightly dependent on the count rate of the source; the photometric aperture and the background value were set according to \citet{SOUSA}. Since the UVOT is a photon-counting detector and suffers from coincidence loss (C-loss) for bright sources, the observed counts need corrections for such losses. This has now been automatically done by the ``uvotsource" program based on an empirical relation given by \citet{UVOTcali}. As the instrumental response curves of the UVOT optical filters do not follow exactly those of the Johnson $UBV$ system, color-term corrections were further applied to the magnitudes \citep{UVOTcali}. Table \ref{Tab:Swiftpho} lists the final UVOT UV/optical magnitudes of SN 2013am.


\begin{deluxetable*}{cccccccc}
\tablewidth{0pt}
\tablecaption{The $UBVRI$ Photometry of SN 2013am from the Ground-based Observations}
\tablehead{\colhead{MJD} & \colhead{Day\tablenotemark{a}} & \colhead{$U$(mag)} & \colhead{$B$(mag)} & \colhead{$V$(mag)} & \colhead{$R$(mag)} & \colhead{$I$(mag)} & \colhead{Telescope} }
\startdata
56373.85 	&	2.35 	&	\nodata	&	16.88(13)	&	16.40(10)	&	15.88(07)	&	15.58(07)	&	LT\tablenotemark{b}	\\
56374.58 	&	3.08 	&	16.69(04)	&	16.90(05)	&	16.41(04)	&	15.86(02)	&	15.48(05)	&	LJT	\\
56375.74 	&	4.24 	&	16.80(08)	&	16.90(07)	&	16.38(06)	&	15.78(07)	&	15.41(05)	&	LJT	\\
56378.87 	&	7.37 	&	\nodata	&	17.08(14)	&	16.29(11)	&	15.67(09)	&	15.31(09)	&	LT\tablenotemark{b}	\\
56379.65 	&	8.15 	&	17.07(07)	&	16.94(06)	&	16.37(03)	&	15.73(03)	&	15.31(07)	&	LJT	\\
56380.96 	&	9.46 	&	\nodata	&	17.07(09)	&	16.31(07)	&	15.64(02)	&	15.24(02)	&	LT\tablenotemark{b}	\\
56381.69 	&	10.19 	&	17.13(06)	&	17.00(06)	&	16.36(04)	&	15.67(07)	&	15.18(07)	&	LJT	\\
56383.73 	&	12.23 	&	17.25(07)	&	17.03(05)	&	16.36(03)	&	15.64(03)	&	15.18(03)	&	LJT	\\
56383.87 	&	12.37 	&	\nodata	&	17.11(12)	&	16.33(10)	&	15.59(07)	&	15.18(07)	&	LT\tablenotemark{b}	\\
56384.73 	&	13.23 	&	17.33(05)	&	17.18(05)	&	16.38(04)	&	15.62(03)	&	15.14(06)	&	LJT	\\
56391.58 	&	20.08 	&	17.96(06)	&	17.50(05)	&	16.41(04)	&	15.61(04)	&	15.08(07)	&	LJT	\\
56394.00 	&	22.50 	&	\nodata	&	17.89(16)	&	16.49(13)	&	15.71(07)	&	15.17(07)	&	LT\tablenotemark{b}	\\
56395.91 	&	24.41 	&	\nodata	&	17.92(09)	&	16.51(08)	&	15.71(05)	&	15.15(05)	&	LT\tablenotemark{b}	\\
56398.55 	&	27.05 	&	\nodata	&	17.70(06)	&	16.48(04)	&	15.65(05)	&	15.03(06)	&	LJT	\\
56399.92 	&	28.42 	&	\nodata	&	17.96(11)	&	16.50(08)	&	15.59(02)	&	14.91(02)	&	LT\tablenotemark{b}	\\
56409.90 	&	38.40 	&	\nodata	&	18.09(09)	&	16.55(07)	&	15.63(04)	&	14.96(04)	&	LT\tablenotemark{b}	\\
56422.54 	&	51.04 	&	\nodata	&	18.39(09)	&	16.51(04)	&	15.63(04)	&	14.89(03)	&	TNT	\\
56422.93 	&	51.43 	&	\nodata	&	18.33(10)	&	16.61(08)	&	15.64(06)	&	14.91(06)	&	LT\tablenotemark{b}	\\
56424.55 	&	53.05 	&	\nodata	&	18.40(14)	&	16.54(08)	&	15.69(07)	&	14.90(03)	&	TNT	\\
56430.88 	&	59.38 	&	\nodata	&	18.46(08)	&	16.67(06)	&	15.64(06)	&	14.87(06)	&	LT\tablenotemark{b}	\\
56437.65 	&	66.15 	&	\nodata	&	18.62(06)	&	16.65(03)	&	15.73(03)	&	14.90(04)	&	LJT	\\
56441.54 	&	70.04 	&	\nodata	&	18.65(11)	&	16.61(07)	&	15.66(04)	&	14.92(05)	&	TNT	\\
56442.54 	&	71.04 	&	\nodata	&	18.71(13)	&	16.63(09)	&	15.60(07)	&	14.90(07)	&	TNT	\\
56444.54 	&	73.04 	&	\nodata	&	\nodata	&	16.65(06)	&	15.65(02)	&	14.89(05)	&	TNT	\\
56446.52 	&	75.02 	&	\nodata	&	\nodata	&	16.67(03)	&	15.66(03)	&	14.94(03)	&	TNT	\\
56455.93 	&	84.43 	&	\nodata	&	18.77(11)	&	16.88(09)	&	15.87(05)	&	15.11(05)	&	LT\tablenotemark{b}	\\
56456.54 	&	85.04 	&	\nodata	&	18.89(11)	&	16.80(06)	&	15.86(06)	&	15.02(05)	&	LJT	\\
56462.91 	&	91.41 	&	\nodata	&	18.85(11)	&	16.93(09)	&	15.94(05)	&	15.21(05)	&	LT\tablenotemark{b}	\\
56480.40 	&	108.90 	&	\nodata	&	\nodata	&	18.30(03)	&	17.22(13)	&	16.41(09)	&	LJT	\\
56588.26 	&	216.76 	&	\nodata	&	\nodata	&	\nodata	&	18.79(07)	&	17.72(06)	&	LT\tablenotemark{b}	\\
56593.26 	&	221.76 	&	\nodata	&	\nodata	&	\nodata	&	18.84(08)	&	17.78(07)	&	LT\tablenotemark{b}	\\
56602.89 	&	231.39 	&	\nodata	&	\nodata	&	20.64(05)	&	19.12(15)	&	17.81(08)	&	LJT	\\
56644.28 	&	272.78 	&	\nodata	&	\nodata	&	\nodata	&	19.15(20)	&	18.19(20)	&	LT\tablenotemark{b}	\\
56653.24 	&	281.74 	&	\nodata	&	\nodata	&	\nodata	&	19.29(10)	&	18.22(10)	&	LT\tablenotemark{b}	\\
56660.95 	&	289.45 	&	\nodata	&	\nodata	&	20.88(05)	&	19.38(06)	&	18.12(06)	&	LJT	\\
56665.85 	&	294.35 	&	\nodata	&	\nodata	&	20.91(13)	&	19.41(08)	&	18.14(07)	&	LJT	\\
56687.94 	&	316.44 	&	\nodata	&	\nodata	&	20.98(10)	&	19.47(10)	&	18.21(08)	&	LJT	\\
56715.74 	&	344.24 	&	\nodata	&	\nodata	&	\nodata	&	19.55(15)	&	18.45(12)	&	LJT
\enddata
\tablecomments{Uncertainties (numbers in brackets), in units of 0.01 mag, are 1$\sigma$; MJD = JD$-2400000.5$.}
\tablenotetext{a}{Relative to the epoch of explosion, Mar. 20.5 2013}
\tablenotetext{b}{Translated from the Sloan $g,r,i$  magnitudes by the relations presented in \citet{ugrizT} and \citet{Jordi06} }
\label{Tab:GDPho}
\end{deluxetable*}

\begin{deluxetable}{ccccc}
\tablewidth{0pt}
\tablecaption{The $gri$ Photometry of SN 2013am from LT}
\tablehead{\colhead{MJD} & \colhead{Day\tablenotemark{a}} & \colhead{$g$(mag)} & \colhead{$r$(mag)} & \colhead{$i$(mag)} }
\startdata
56373.85	&	2.35	&	16.70(07)	&	16.14(05)	&	16.02(03)	\\
56378.87	&	7.37	&	16.70(09)	&	15.94(02)	&	15.79(07)	\\
56380.96	&	9.46	&	16.70(06)	&	15.97(01)	&	15.78(01)	\\
56383.87	&	12.37	&	16.71(07)	&	15.90(04)	&	15.70(04)	\\
56394.00	&	22.50	&	17.11(10)	&	16.07(04)	&	15.74(04)	\\
56395.91	&	24.41	&	17.14(06)	&	16.08(02)	&	15.73(04)	\\
56399.92	&	28.42	&	17.16(07)	&	16.06(01)	&	15.60(01)	\\
56409.90	&	38.40	&	17.26(05)	&	16.07(03)	&	15.61(02)	\\
56422.93	&	51.43	&	17.43(06)	&	16.05(03)	&	15.54(04)	\\
56430.88	&	59.38	&	17.53(05)	&	16.07(01)	&	15.51(05)	\\
56455.93	&	84.43	&	17.80(07)	&	16.25(03)	&	15.70(03)	\\
56462.91	&	91.41	&	17.87(06)	&	16.28(04)	&	15.76(02)	\\
56588.26	&	216.76	&	\nodata	&	18.78(04)	&	18.28(07)	\\
56593.26	&	221.76	&	\nodata	&	18.84(06)	&	18.30(04)	\\
56644.28	&	272.78	&\nodata	&	19.26(08)	&	18.62(07)	\\	
56653.24	&	281.74	&	\nodata	&	19.43(06)	&	18.68(06)
\enddata
\tablecomments{Uncertainties (numbers in brackets), in units of 0.01 mag, are 1$\sigma$; MJD = JD-2400000.5.}
\tablenotetext{a}{Relative to the date of explosion, Mar. 20.5 2013}
\label{Tab:gri}
\end{deluxetable}

\begin{deluxetable*}{cccccccc}[!b]
\tablewidth{0pt}
\tablecaption{$Swift$ UVOT Photometry of SN 2013am}
\tablehead{\colhead{MJD} & \colhead{Day\tablenotemark{a}} & \colhead{uvw2(mag)} & \colhead{uvm2(mag)} & \colhead{uvw1(mag)} & \colhead{$U$(mag)} & \colhead{$B$(mag)} & \colhead{$V$(mag)} }
\startdata
56373.82	&	2.32	&	17.80(20)	&	18.30(20)	&	17.30(20)	&	16.70(20)	&	\nodata	&	\nodata	\\
56378.86	&	7.36	&	19.00(15)	&	19.40(20)	&	17.90(20)	&	17.89(18)	&	16.99(12)	&	16.38(0.07)	\\
56380.26	&	8.75	&	19.26(18)	&	\nodata	&	18.29(27)	&	17.95(17)	&	17.04(09)	&	16.36(0.06)	\\
56380.63	&	9.13	&	19.40(13)	&	20.03(12)	&	18.48(31)	&	18.03(20)	&	17.03(10)	&	16.37(0.05)	\\
56382.42	&	10.92	&	19.71(11)	&	20.16(15)	&	18.83(25)	&	18.12(15)	&	17.05(10)	&	16.36(0.04)	\\
56384.46	&	12.96	&	19.53(22)	&	20.72(20)	&	19.09(30)	&	18.25(21)	&	17.03(09)	&	16.38(0.06)	\\
56386.42	&	14.92	&	19.85(17)	&	\nodata	&	\nodata	&	18.31(15)	&	17.24(10)	&	16.39(0.06)	\\
56386.76	&	15.26	&	19.71(17)	&	\nodata	&	\nodata	&	18.43(20)	&	17.28(09)	&	16.40(0.05)	\\
56388.42	&	16.92	&	\nodata	&	\nodata	&	\nodata	&	18.65(13)	&	17.33(08)	&	16.38(0.05)	\\
56390.39	&	18.89	&	19.96(21)	&	\nodata	&	\nodata	&	18.78(16)	&	17.48(10)	&	16.35(0.04)	\\
56400.69	&	29.18	&	\nodata	&	\nodata	&	\nodata	&	19.50(17)	&	17.78(08)	&	16.39(0.05)
\enddata
\tablecomments{Uncertainties (numbers in brackets), in units of 0.01 mag, are 1$\sigma$; MJD = JD$-2400000.5$.}
\tablenotetext{a}{Relative to the date of explosion, Mar. 20.5 2013}
\label{Tab:Swiftpho}
\end{deluxetable*}


\subsection{Spectroscopy}
\label{subsect:sp}
A total of 11 spectra were obtained with the YFOSC and LJT, covering the wavelength from $\sim$3400\AA~to $\sim$9100\AA (with a resolution of $\sim$18\AA~).  A spectrum taken with the low resolution spectrograph on the 3.58-m Telescopio Nazionale Galileo (hereinafter TNG) at La Palma \citep{ATel4910} is also included in our analysis. These spectra were reduced using standard IRAF routines and were flux-calibrated with  spectrophotometric flux standard stars observed at a similar air mass on the same night. The continuum of the spectra were adjusted to match the $BVRI$ photometry with a precision of 0.1 mag.  The spectra were further corrected for the atmospheric absorption at the observing site, and telluric lines were also removed from the data. A journal of the spectroscopic observations is given in Table \ref{Tab:Spec_log}.

\begin{figure}
\centering
\includegraphics[width=8cm,angle=0]{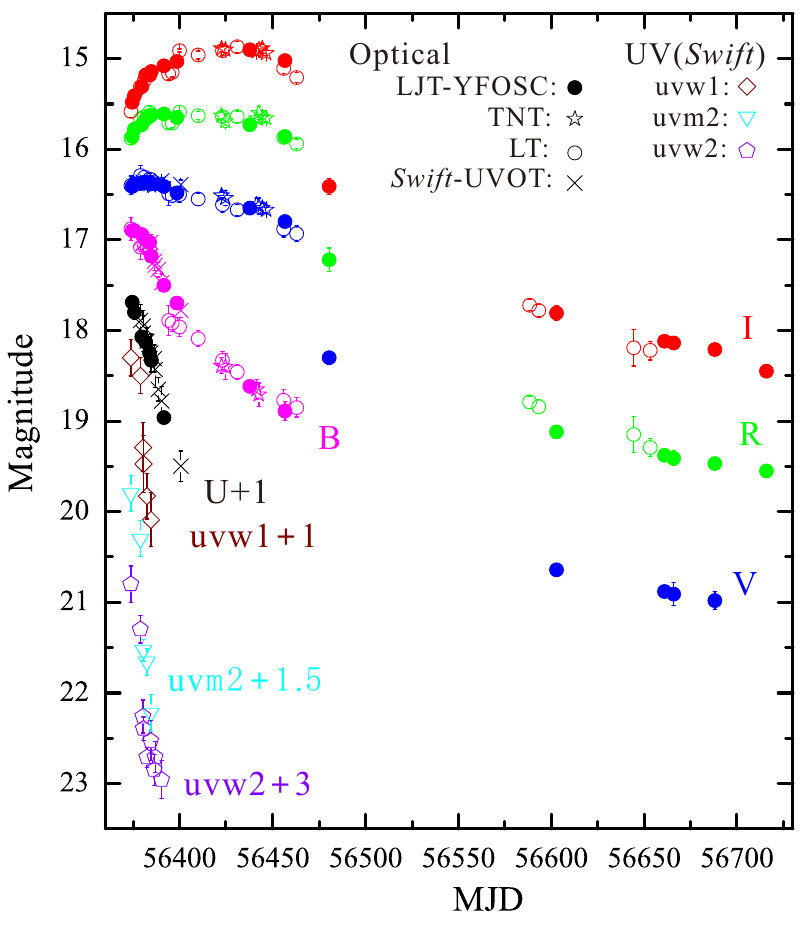}
 \caption{UV and optical light curves of SN 2013am, which are shifted vertically for better display.}
 \label{<LC>}
\end{figure}
\begin{deluxetable*}{cccccccc}
\tablewidth{0pt}
\tablecaption{Journal of Spectroscopic Observations of SN 2013am}
\tablehead{\colhead{UT Date}&\colhead{MJD} & \colhead{Epoch\tablenotemark{a}} & \colhead{Res} & \colhead{Range}  & \colhead{Airmass} & \colhead{Exp.time}& \colhead{Telescope} \\
\colhead{}&\colhead{(-240000.5)} & \colhead{(days)} & \colhead{(\AA/pix)} & \colhead{(\AA)} & \colhead{} & \colhead{(sec)} &\colhead{(+Instrument)}  }
\startdata
Mar.	23.1	&	56374.10	&	2.60	&	2.72	&	3340-8080	&	1.18	&	600.00	&	TNG+LRS	\\
Mar.	23.6	&	56374.58	&	3.08	&	2.87	&	3500-9100	&	1.22	&	1200.00	&	LJT+YFOSC	\\
Mar.	24.8	&	56375.74	&	4.24	&	2.87	&	3490-9100	&	1.10	&	1200.00	&	LJT+YFOSC	\\
Mar.	28.8	&	56379.65	&	8.15	&	2.87	&	3510-9120	&	1.19	&	1200.00	&	LJT+YFOSC	\\
Mar.	30.7	&	56381.69	&	10.19	&	2.87	&	3520-9080	&	1.06	&	1200.00	&	LJT+YFOSC	\\
Apr.	01.8	&	56383.73	&	12.23	&	2.87	&	3480-9120	&	1.15	&	1200.00	&	LJT+YFOSC	\\
Apr.	02.8	&	56384.73	&	13.23	&	2.87	&	3510-9100&	1.21	&	1200.00	&	LJT+YFOSC	\\	
Apr.	09.6	&	56391.58	&	20.08	&	2.87	&3490-9120	&	1.07	&	1200.00	&	LJT+YFOSC	\\	
Apr.	16.6	&	56398.55	&	27.05	&	2.87	&3520-9150	&	1.10	&	1200.00	&	LJT+YFOSC	\\	
Apr.	22.6	&	56404.63	&	33.13	&	2.87	&3510-9140	&	1.04	&	1200.00	&	LJT+YFOSC	\\	
May	25.7	&	56437.65	&	66.15	&	2.87	&3510-9100	&	1.13	&	1200.00	&	LJT+YFOSC	\\	
Jun.	11.5	&	56454.54	&	83.04	&	2.87	&3520-9080	&	1.14	&	1200.00	&	~~~~LJT+YFOSC		
\tablecomments{Journal of spectroscopic observations of SN 2013am.}
\tablenotetext{a}{Relative to the date of explosion, Mar. 20.5 2013}
\enddata

\label{Tab:Spec_log}
\end{deluxetable*}

\section{Light Curves of SN 2013am}
\label{sect:LC}

Figure \ref{<LC>} shows the optical and UV light curves of SN 2013am. The light curves in the UV bands show a rapid evolution compared to those in the optical bands. The post-maximum magnitude declined at a rate of 10-30 mag $(100d)^{-1}$ in UV, while it remains nearly constant in the optical. Detailed analysis of the light and color curves are presented in the following subsections.

\subsection{Optical Light Curves}
\label{subsect:OpLC}
\begin{figure*}
\centering
\includegraphics[width=13cm,angle=0]{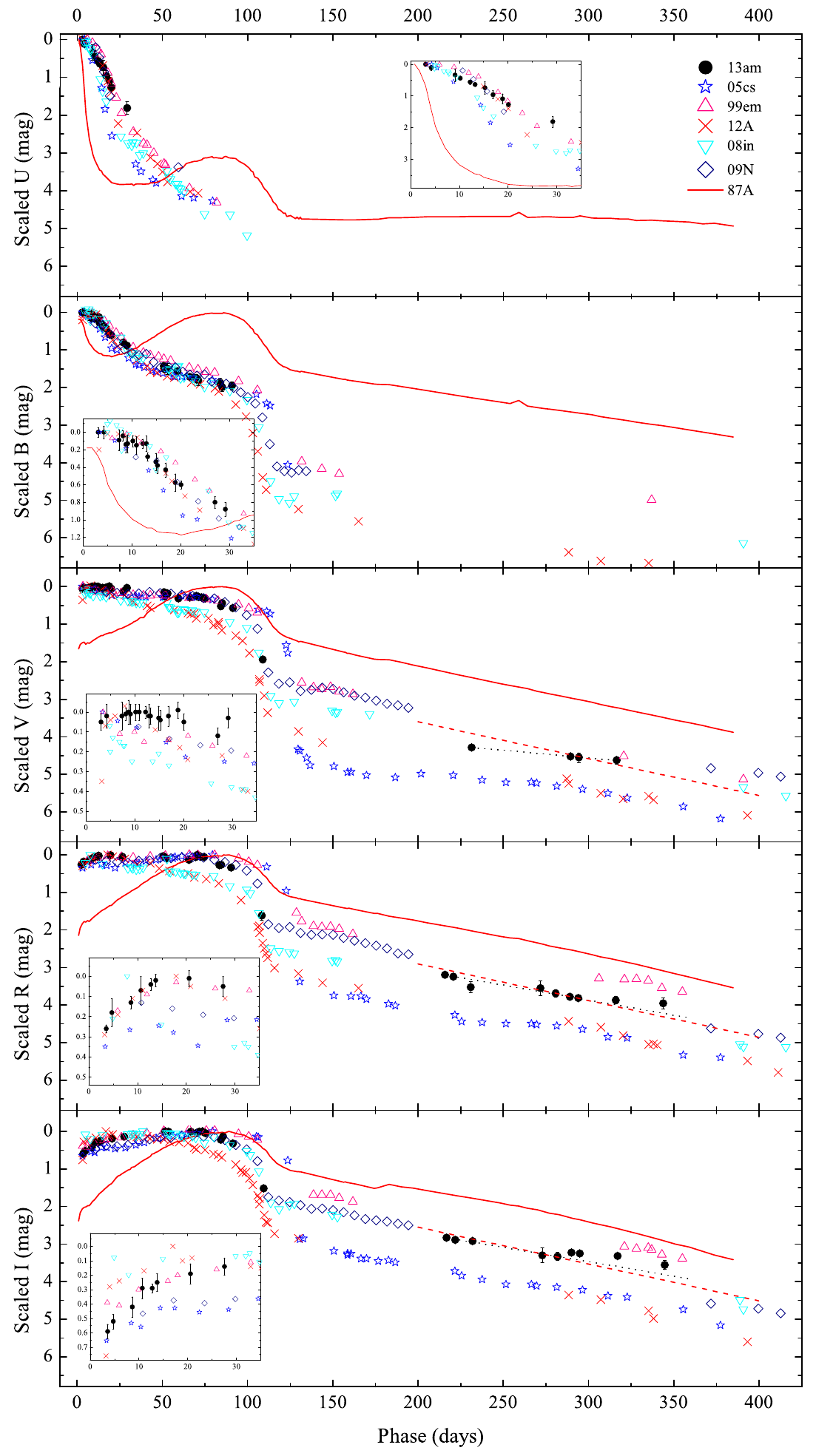}
 \caption{$UBVRI$ light curve comparisons of SN 2013am. The well-observed SNe IIP 2005cs, 1999em, 2008in, 2009N,  2012A and 1987A are overplotted for comparison. The dashed lines in panel $V$, $R$, and $I$ represent the slope of the decay from Co to Fe. All the light curves were normalized to the peak. The inset in each panel shows the light curves at early phases. See text for the references of the comparison SNe.}
  \label{<LCcomp>}
\end{figure*}

The optical light curves of SN\,2013am are shown in Figure \ref{<LCcomp>}. Overplotted are those of representative SNe IIP, including the normal objects SNe 1999em (\citealp{Leo_99em, Ham_99em,Elm_99em}) and 2012A \citep{SN12A}, the low-velocity one SN 2005cs (\citealp{05csa, 05csb}), and the `gap-filler' events SNe 2008in \citep{Roy11} and 2009N \citep{2009N}. The light curves of peculiar SN IIP 1987A \citep{87A1,87A2,87A3,87A4} are also shown for comparison. Applying a polynomial fit to the observed data, we found that SN 2013am reached at the $B$- and $V$- band maximum at $\sim$+2 and $\sim$+7 days after the explosion, with $m_B =16.90\pm0.05$ mag and $m_V = 16.36\pm0.03$ mag, respectively.

In the $U$ band, SN 2013am shows a post-maximum evolution similar to SN 2009N and SN 1999em, with a magnitude decline of about 1.2 mag in 20 days from the maximum light. In comparison, SN 2005cs and SN 2008in have a magnitude decline of $\sim$2.5 mag and $\sim$2.0 mag, respectively, at similar phases. This difference seen in the $U$ band becomes marginally important in other bands. However, the light curves of different SNe IIP show significant scatter in the transition and radioactive-decay phases. For example, the magnitude decline from the end of the plateau to the tail phase (i.e., at $t\sim$ +150 days after explosion) can vary from about 2.0 mag (for SN 1999em) to over 4.0 mag (for SN 2005cs) in the $V$ band.  Although SN 2013am and SN 2005cs have very similar spectroscopic features (see discussions in section \ref{subsect:SpEv}), the former has a tail that is $\sim$1.0 mag brighter than the latter. A brighter tail indicates that more nickel is likely produced in the explosion. One can see that SN 2013am bears a close resemblance to the transition object SN 2009N in light of the entire shape of the light curves. Although SN 2013am also shows some similarities to SN 1999em, but it has a shorter duration of ``plateau" phase.

To examine the evolution of the late-time light curves, we also measured the decline rates in $VRI$ bands during the period $\sim$220-320 days (Figure \ref{<LCcomp>}).  For SN 2013am, we obtained $\gamma_V$ = 0.42$\pm0.02$, $\gamma_R$ = 0.59$\pm$0.07, and $\gamma_I$ = 0.54$\pm0.06$ mag/$100^d$. These values are smaller than that of SN 2005cs derived at similar phases (i.e., $\gamma_V$ = 0.65$\pm0.06$, $\gamma_R$ = 0.68$\pm$0.06, and $\gamma_I$ = 0.78$\pm0.06$ mag/$100^d$, derived from the light curves of SN 2005cs in \citealp{05csb}).  Note that the decay rates derived for these two SNe IIP are smaller than that expected from the Co$\rightarrow$Fe decay (i.e., $\gamma = 0.98$ mag/$100^d$) especially in the $V$ band. This phenomenon was also observed in low-luminous, low-energy events \citep{05csb}. It implies that the tail luminosity of SN 2013am and SN 2005cs do not fall directly onto the radioactive tail, which might be due to contribution from additional radiation generated in the warmer inner ejecta propagating throughout the transparent cooler external layers \citep{Utrobin}. The deviation of the last two points in $R$- and $I$- band perhaps arises from the photometric errors due to the contamination by host-galaxy light. Excluding the last two points, the decay rates become as $\gamma_R$ = 0.79$\pm$0.02 and $\gamma_I$ = 0.79$\pm0.04$ mag/$100^d$.

\subsection{UV Light Curves}
\label{subsect:UVLC}

\begin{figure}
\centering
\includegraphics[width=8.2cm,angle=0]{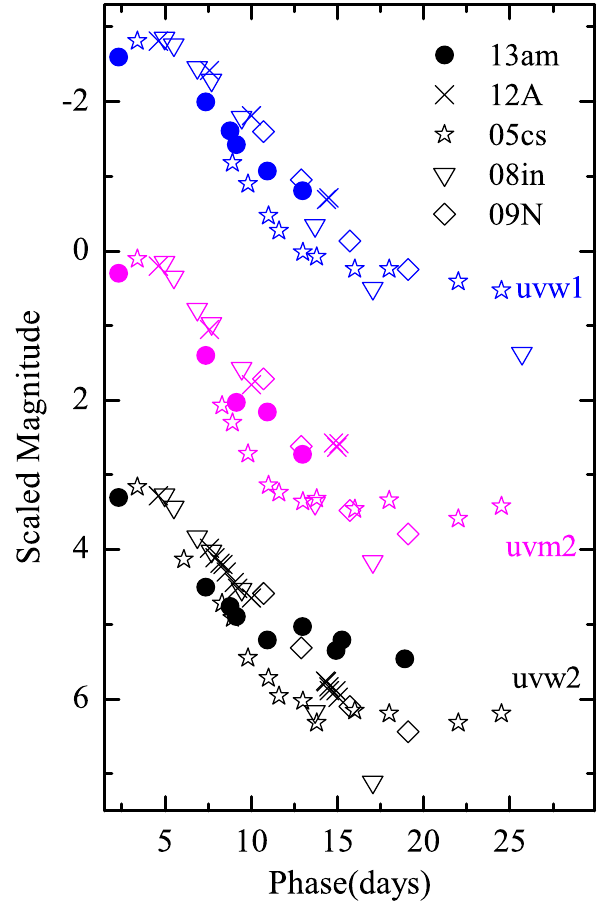}
\caption{A comparison of $Swift$ UV light curves of SN 2013am with SNe, 2005cs,
2008in, 2009N and 2012A. All the light curves were normalised to the peak
 and in uvw1 and uvw2 bands were shifted by $-3$ and $+3$ mag,
respectively. See text for details.}
  \label{<UVLC>}
\end{figure}

The $Swift$ UV light curves of SN 2013am are presented in Figure \ref{<UVLC>}, together with those of SN 2005cs \citep{Browncs}, SN 2008in \citep{Roy11}, SN 2009N \citep{2009N}, and SN 2012A \citep{Prit}. In the uvw2 band, SN 2013am shows a somewhat flatter evolution than SN 2005cs; and it is unclear whether this trend holds in the uvm2 and uvm1 bands for the lack of data. After $t\sim$10 days from the explosion, SN 2013am seems to show a relatively higher luminosity than the comparison SNe.

\subsection{Color Curves}
\label{subsect:CC}

\begin{figure*}
\centering
\includegraphics[width=17cm,angle=0]{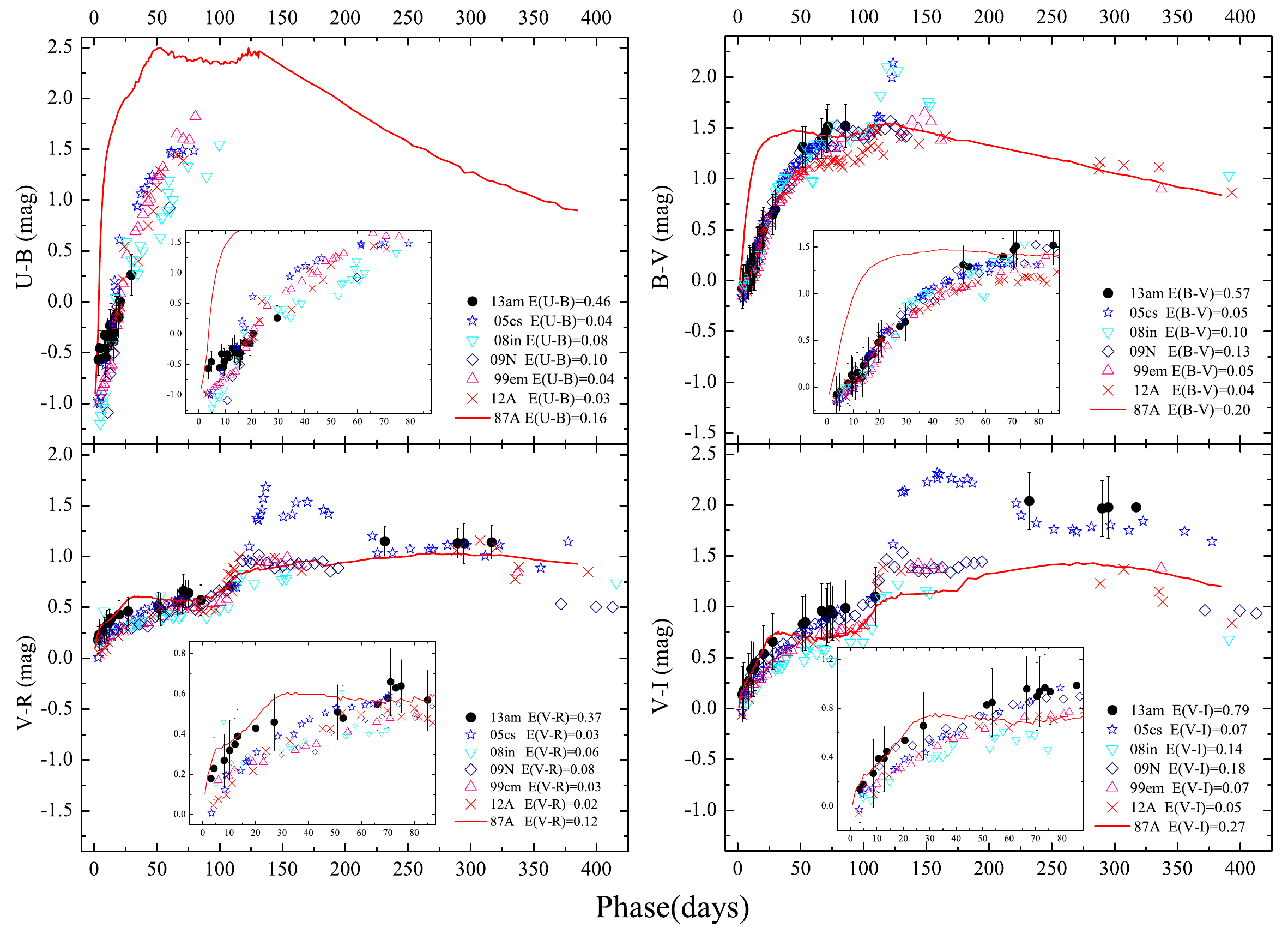}
\caption{Comparison of $UBVRI$ color curves between SN 2013am and some representative SNe IIP: 1987A, 1999em, 2005cs, 2008in, 2009N, and 2012A. All the color curves were de-reddened by the corresponding color excess shown in each panel. The errors for SN 2013am include the uncertainties in photometry and reddening corrections. The insets show the evolution of the light curves at early time.}
\label{<CC>}
\end{figure*}

Figure \ref{<CC>} shows the color curves of SN 2013am, corrected for the Galactic ($E(B-V)_{MW}$=0.02) and host-galaxy reddening ($E(B-V)_{host}$=0.55$\pm$0.19 mag) that is derived in section \ref{subsect:ExtMag}).  The color curves of the comparison SNe are also plotted. We found that the color curves of SN 2013am show an evolution that is similar to the comparison SNe IIP during the plateau phase. This similarity of early color curves has been also noticed for a large SN IIP sample (i.e., Faran et al. 2014).

It is notable that the color of SNe IIP evolve towards red more quickly in bluer bands at the early phase owing to that the luminosity at shorter wavelengths is decreasing more quickly. In the nebular phase, the $V-I$ color is found to be unusually red in SN 2013am and SN 2005cs. On the other hand, \citet{Faran14} found that the decline rate of the $I$-band light curve correlates with the ejecta velocity at $t\approx50$ days after explosion. They proposed that high ejecta velocities may indicate a fast decrease in density, which results in a quicker release of radiation. The released radiation would heat the SN for a while, which consequently slows down the decrease of the photospheric temperature of some SNe IIP. In other words, the temperature of SNe IIP with slower ejecta velocity tends to be cooling quicker. Therefore, the unusual red $V-I$ color of faint SNe IIP might be a typical feature of low-velocity (or low-luminosity) SNe IIP.

\subsection{Extinction}
\label{subsect:ExtMag}
\citet{Ham03} suggested using the $V - I$ color as a better reddening indicator because it is expected to be less sensitive to metallicity effects.
\citet{Kri09} found that the $V-[RIJHK]$ color curves of SNe IIP tend to evolve in a similar way during the plateau phase, which might allow us to determine the
host-galaxy reddening that SN 2013am suffered by comparing with the SN sample shown in Figure \ref{<CC>}. As we do not have the near infrared photometry of SN 2013am, we apply this method to the $V - I$ color curve and derive a reddening of E$(B - V)$ = 0.65$\pm$0.08 mag for SN 2013am by comparing with the un-reddened $V - I$ template from Krisciunas et al. (2009). Furthermore, \citet{Oli10} suggest that the host reddening can be derived with a precision of $\sigma(A_V)$ = 0.2 mag from the $V - I$ color toward the end of the plateau, which follows as:
\begin{equation}
A_V(V - I)=\beta_V[(V - I)-0.656],
\end{equation}
where the $V - I$ color corresponds to the color of a given SN at 30 days before the end of plateau, corrected for K-terms and Galactic reddening; and $\beta_V = A_V/E(V - I)$ is related with $R_V$ (e.g., $\beta_V$=2.518 for $R_V$=3.1; \citealp{Oli10}). Based on the $V - I$ color of SN 2013am measured at the appropriate phase, the host-galaxy reddening is estimated as $E(B - V)_{host}$=0.54$\pm$0.15 mag for $R_V$ = 3.1 and $E(B - V)_{host}$=0.64$\pm$0.12 mag for $R_V$=1.4.

\citet{Nug06} assumed a correlation between the $I$-band luminosity and velocity of \ion{Fe}{2} (in km s$^{-1}$) at $t\approx+50$ days:
\begin{equation}
M_I =M_{I_0}-\alpha log(v_{FeII}/5000)-R_I[(V - I)-(V - I)_0],
\end{equation}
where $M_{I_0}$=17.49$\pm$0.08 mag, $\alpha$=6.69$\pm0.5$, $(V-I)_0$= 0.53, and $R_I$=1.36 (corresponding to R$_{V}$ = 3.1, \citealp{Nug06}). Based on an expanded sample of SNe IIP, \citet{Ext09} allowed R$_{I}$ to be a free parameter and obtained $M_{I_0}$=17.43$\pm$0.10 mag, $\alpha$=4.6$\pm0.7$, and $R_I=0.7^{+0.3}_{-0.4}$ (corresponding to $R_{V}$ = 1.5$\pm0.5$). However, these two methods yield a similar estimate of the extinction $A_V$$\sim$1.0 mag (see also Table \ref{Tab:Extin}) under the assumption that SN 2013am follows the velocity-luminosity relation. However, this assumption may not be valid (see discussions in section \ref{subsect:Diver}). Thus, the reddening derived from the above empirical correlation are not adopted in our analysis. The values of A$_{V}$ determined from the velocity-luminosity relation for R$_{V}$ = 3.1 and 1.4 were listed in table 6.

Besides the $V - I$ color curve, the absorption strength of Na I lines in the spectra provides an alternative diagnostic of the host-galaxy reddening.
In SN 2013am, an obvious \ion{Na}{1} D absorption feature can be detected in the early spectra of SN 2013am, as shown in Figure \ref{<TriL>}. The equivalent width ($EW$) of the \ion{Na}{1} D absorption measured from our low-resolution spectra is 1.65$\pm$0.08\AA.  Several empirical correlations between reddening and $EW$ of \ion{Na}{1} D have been applied to estimate the reddening as listed in Table \ref{Tab:Extin}, which yield a mean value of $E(B-V)_{host}$ = 0.52 $\pm$0.24 mag. The large error indicates that the correlation of reddening with the absorption strength of Na I D lines is not very reliable \citep{Ext11}. Combining the results from the photometric and spectroscopic methods, we obtain the host-galaxy reddening as $E(B-V)_{host}$ = 0.55$\pm$0.19 mag for SN 2013am.

\begin{deluxetable*}{cccc}
\tablecaption{ The Host-galaxy Reddening Derived for  SN 2013am}
\tablehead{\colhead{Method } &\colhead{$A_V$(mag)} & \colhead{Formula} & \colhead{Results(mag)} }
\startdata
Color Curve &1.95$\pm$0.25&$E(B-V)=E(V-I)/1.39$\tablenotemark{a} & 0.63$\pm$0.08 \\
Color Curve &1.45$\pm$0.20&$E(B-V)=E(V-I)/1.19$\tablenotemark{b} & 0.76$\pm$0.14 \\
Color Curve &1.35$\pm$0.59&Formula 1\tablenotemark{c} & 0.44$\pm$0.19 \\
Color Curve &1.01$\pm$0.27&Formula 1\tablenotemark{d} & 0.51$\pm$0.19 \\
Velocity-Luminosity &1.05$\pm$0.37&$E(B-V)=A_V/3.1$\tablenotemark{e} & 0.34$\pm$0.12 \\
VelocityLuminosity &1.41$\pm$0.45&$E(B-V)=A_V/1.5$\tablenotemark{f} & 0.94$\pm$0.30 \\
\ion{Na}{1} D &\nodata\tablenotemark{j}&$E(B-V)=0.25EW$\tablenotemark{g} & 0.41$\pm$0.20  \\
\ion{Na}{1} D &\nodata\tablenotemark{j}&$E(B-V)=0.16EW-0.01$\tablenotemark{h} & 0.25$\pm$0.20\\
\ion{Na}{1} D &\nodata\tablenotemark{j}&$E(B-V)=0.51EW-0.04$\tablenotemark{h} & 0.80$\pm$0.20\\
\ion{Na}{1} D &\nodata\tablenotemark{j}&$E(B-V)=0.43EW-0.08$\tablenotemark{i} &0.63 $\pm$0.30\\
\tablenotetext{a}{$R_V$=3.1, refers to the $V-I$ color of comparison SNe in Figure \ref{<CC>}.}
\tablenotetext{b}{$R_V$=1.4, refers to the $V-I$ color of comparison SNe in Figure \ref{<CC>}.}
\tablenotetext{c} {$R_V$=3.1, \citet{Oli10}}
\tablenotetext{d} {$R_V$=1.4, \citet{Oli10}}
\tablenotetext{e} {$A_V$ was derived from Formula 2 in \citet{Nug06}}
\tablenotetext{f} {$A_V$ was derived from Formula 2 with updated parameter in \citet{Ext09}}
\tablenotetext{g} {\citet{Bar25} }
\tablenotetext{h} {\citet{Tura16}}
\tablenotetext{i} {\citet{Ext11}}
\tablenotetext{j} {Independent to reddening law.}
\enddata
\label{Tab:Extin}
\end{deluxetable*}

\section{Spectroscopy}
\label{sect:Spe}

\begin{figure*}[!th]
\centering
\includegraphics[width=15cm,angle=0]{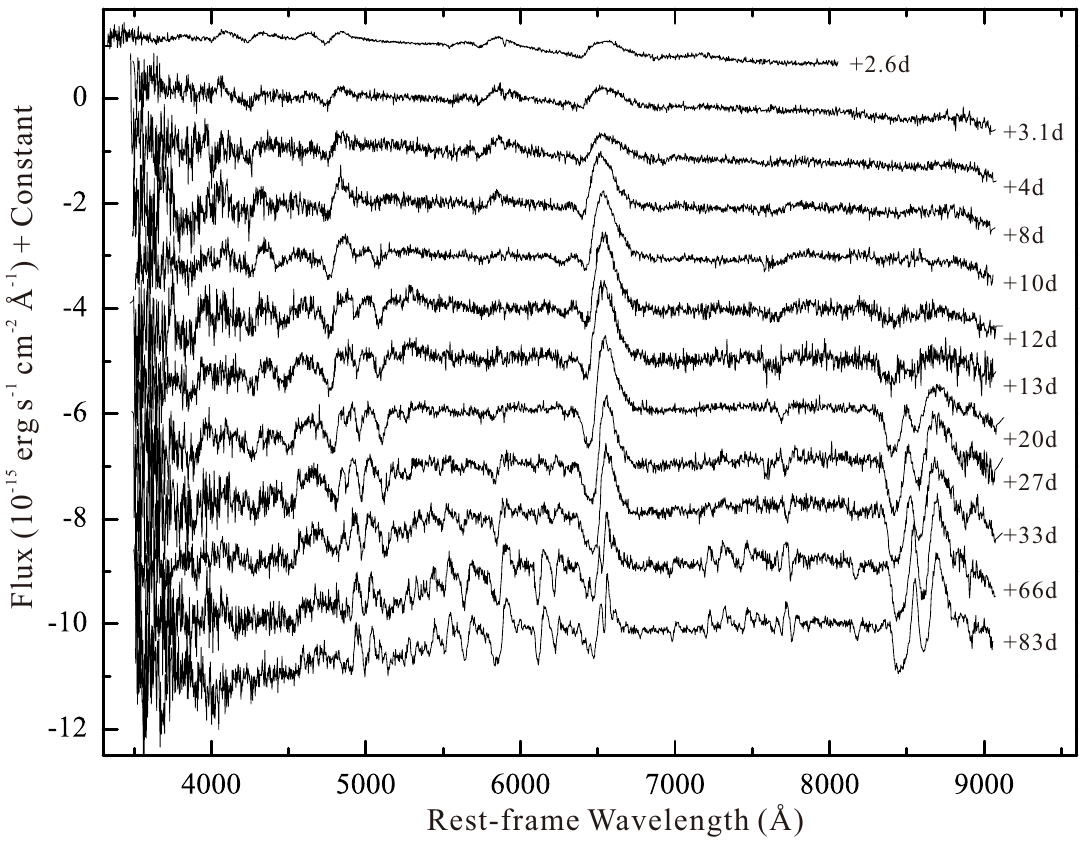}
\caption{Optical spectral evolution of SN 2013am. The spectra have been corrected for the redshift of the host galaxy ($V_{hel}$ = 807 km s$^{-1}$) and telluric absorptions. They have been shifted vertically for better display. The numbers on the right side mark the epochs of the spectra in days after explosion. }
  \label{<WhSp>}
\end{figure*}

A total of 12 spectra of SN 2013am taken by TNG and LJT are plotted in Figure \ref{<WhSp>}, which covers phases from $t \approx +2$ to $t \approx +83$ days
after explosion. The early-time spectra are characterized by asymmetric H$\alpha$ and H$\beta$ emission with prominent P-Cygni profile. A weak \ion{He}{1} $\lambda$5876 emission is also visible in the earlier spectra. The red continuum suggests significant line-of-sight reddening towards SN 2013am.

\subsection{Temporal Evolution of the Spectra}
\label{subsect:SpEv}
\begin{figure*}[!th]
\centering
\includegraphics[width=12cm,angle=0]{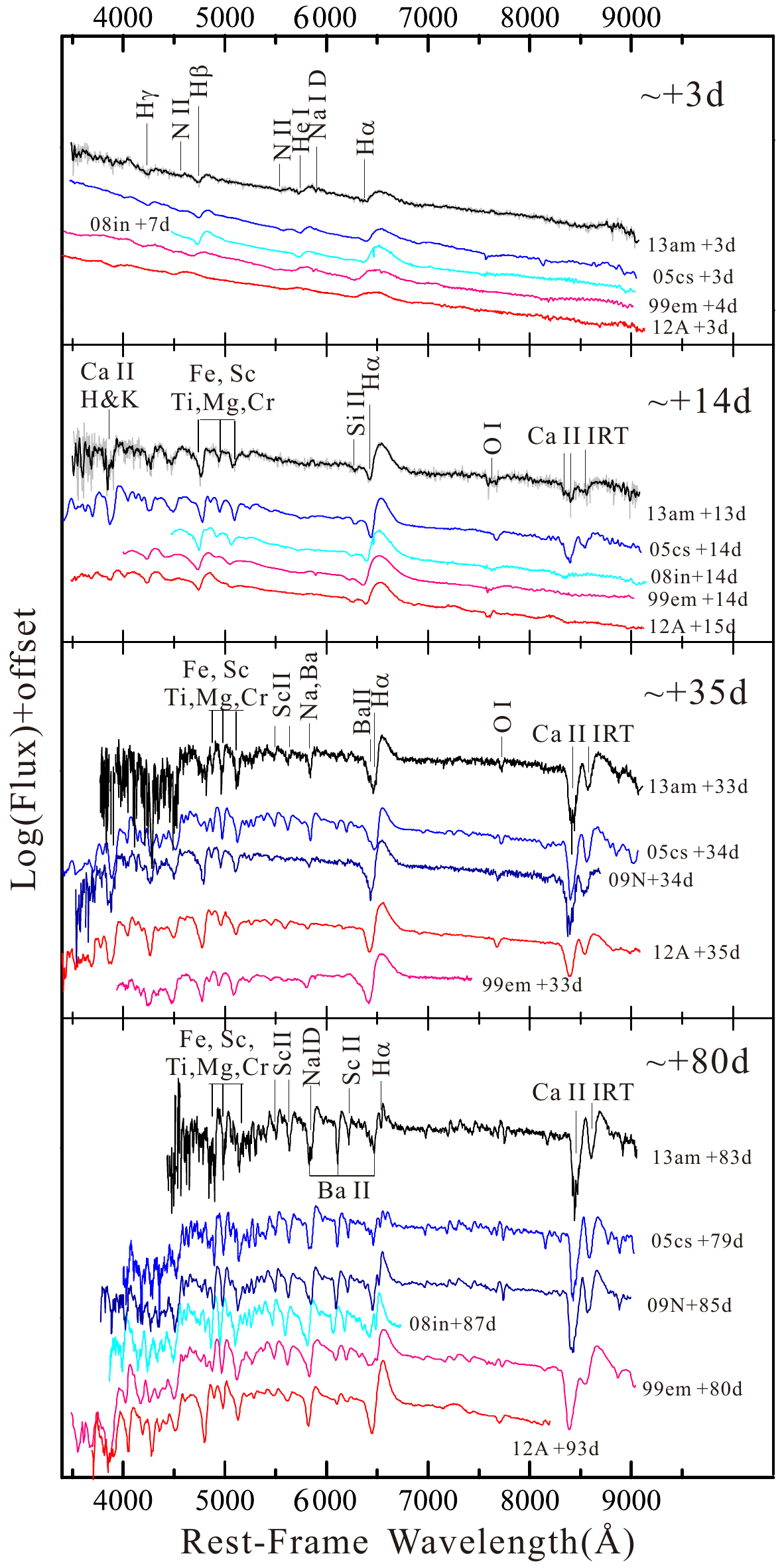}
\caption{Spectra comparison of SN 2013am with SNe 2005cs, 2008in, 2009N, 2012A, and 1999em at four selected epochs. The spectra of SN 2013am at $t \approx $+3 and $t \approx$ +13 days were smoothed with a resolution of 30\AA for a better display. The main spectral features are identified following the spectra of SN 2005cs \citep{05csb}.}
\label{<Spcomp>}
\end{figure*}

Spectral comparisons between SN 2013am and five well-observed SNe IIP at four selected epochs are shown in Figure \ref{<Spcomp>}.  All spectra have been
corrected for redshift and reddening.

At $t \approx +3$ days, all of the comparison SNe show very blue continuum, with dominant line features of H Balmer and \ion{He}{1} $\lambda$5876. The minor absorption feature on the blue side of \ion{He}{1} $\lambda$5876 is likely due to \ion{N}{2} $\lambda$5679. This feature is similarly identified in SN 2005cs \citep{05csa} and might relate to nitrogen enrichment \citep{Hill98}. A relatively strong absorption in the spectrum of SN 2013am near 5900\AA~is due to the host \ion{Na}{1} D, which implies a significant line-of-sight reddening.

A notable change in the $t \approx +14$ days spectrum is that the H$\alpha$ absorption at $\sim$ 6400\AA\ is well developed. The minor absorption on the
blue wing can be attributed to \ion{Si}{2} $\lambda$6355 rather than high velocity feature of H$\alpha$ \citep{05csa}. At this phase, only SN 2013am and SN 2005cs show strong \ion{Ca}{2} features (H$\&$K and IR triplet). Whereas this feature is not detectable in normal SNe IIP by $t\sim$2 weeks after explosion. The fact that \ion{Ca}{2} absorptions are only detected in SN 2013am and SN 2005cs at this phase may indicate a larger Ca/O ratio for their progenitors, which in turn points to a smaller progenitor mass because the Ca/O ratio is expected to decrease with the progenitor mass \citep{HouckFransson}.

\begin{figure}[tb]
\centering
\includegraphics[width=8cm,angle=0]{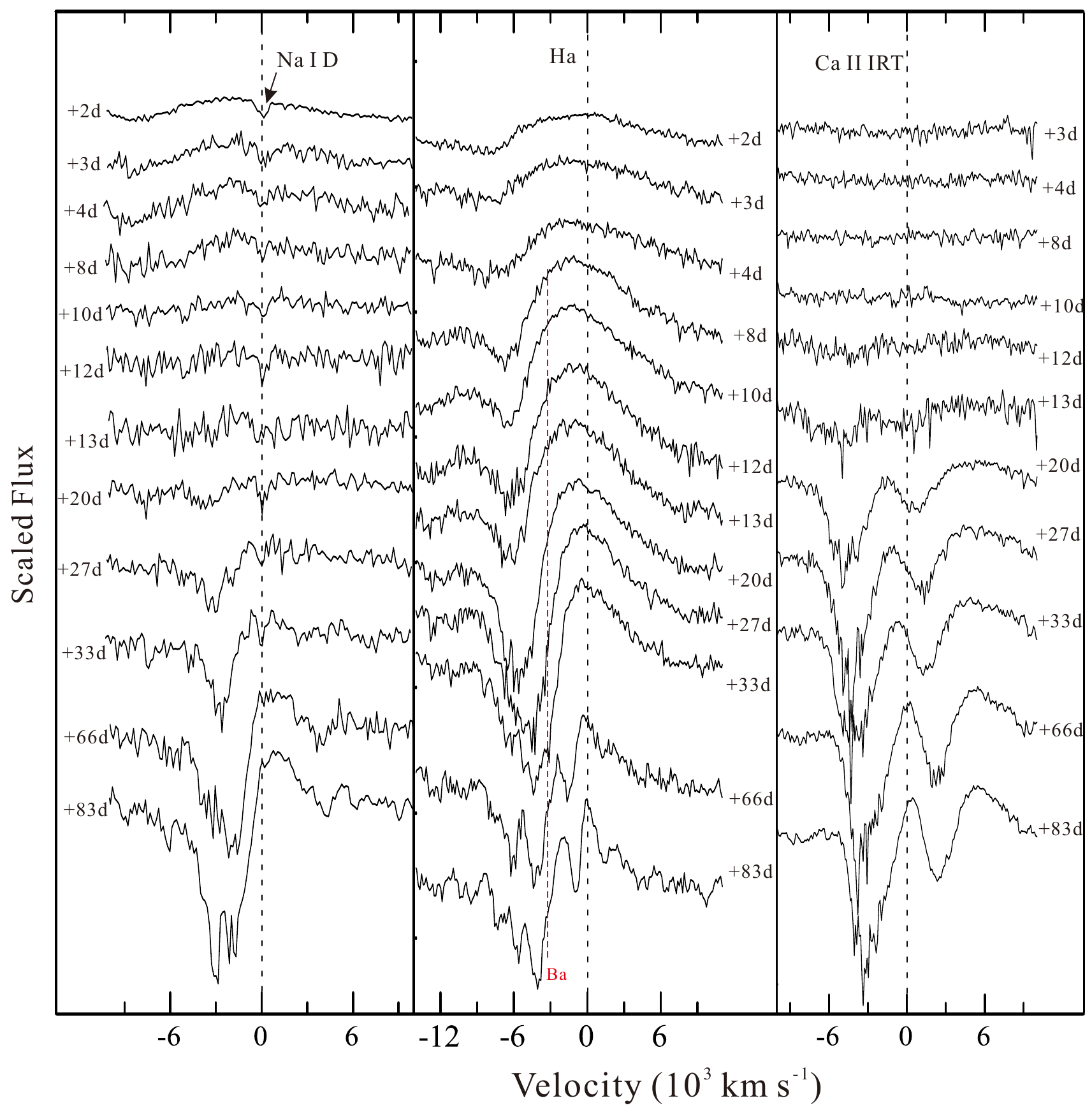}
 \caption{Evolution of selected spectral features of SN 2013am during the photospheric phases. Short dashed lines mark the rest wavelength positions of  \ion{Na}{1} D (5893\AA), H$\alpha$ (6563\AA) and \ion{Ca}{2} IR triplet (8542\AA); and the red dashed line is for \ion{Ba}{2} $\lambda$6497.}
 \label{<TriL>}
\end{figure}

\begin{figure}[!th]
\centering
\includegraphics[width=8cm,angle=0]{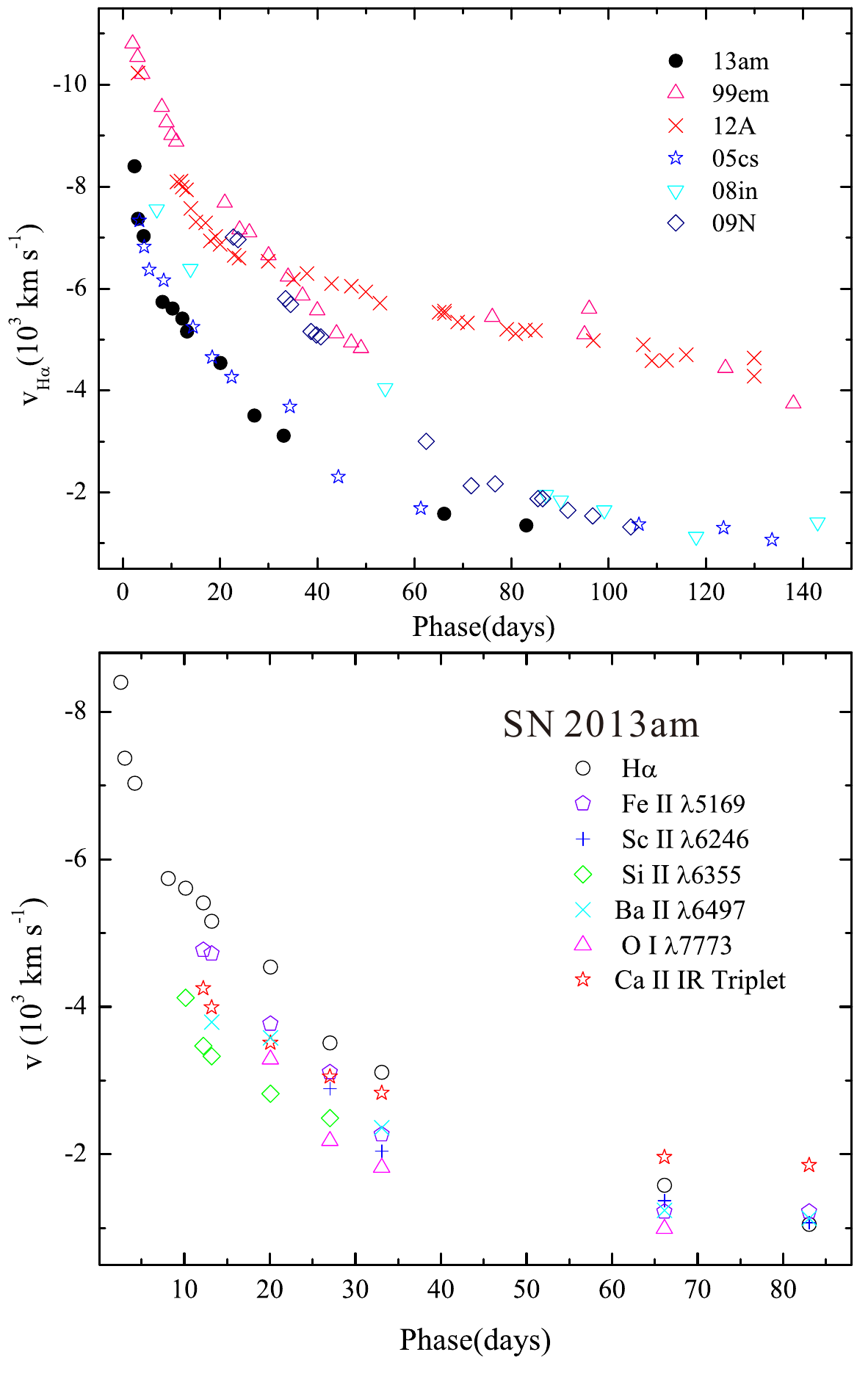}
 \caption{Top panel: the evolution of the line velocities of H$\alpha$ for SN 2013am, compared with those of SNe 2005cs,  1999em,  2008in, 2009N, and 2012A. Bottom panel: the evolution of ejecta velocities of SN 2013am inferred from \ion{Fe}{1} $\lambda$5169, \ion{Sc}{2} $\lambda$6246, \ion{Si}{2} $\lambda$6355,  H$\alpha$, \ion{O}{1} $\lambda$7773 and \ion{Ca}{2} $\lambda$8662 in the spectra. }
 \label{<VHa>}
\end{figure}

By $t \sim 1$ month, the supernova enters the plateau phase and the spectrum becomes red as the decrease of the photospheric temperature. The absorption feature of Ca II IR triplet is well developed. In particular, numerous narrow metal lines (\ion{Fe}{2} , \ion{Ti}{2}, \ion{Sc}{2}, \ion{Ba}{2}, \ion{Cr}{2}, \ion{Sr}{2} and \ion{Mg}{2}) emerge to dominate the spectrum (see \citealp{Past04} for a detailed line identification).

At $t\approx$+80 days, the supernova enters near the end of the plateau phase when the continuum spectrum becomes progressively redder. The P-Cygni profile of H lines becomes much narrower at this phase, reaching at an expansion velocity of about 1000 km s$^{-1}$. The narrow metal lines become more noticeable in SN 2013am and other comparison objects. In particular, the prominent absorption features of \ion{Ba}{2} lines are visible in SN 2013am at $\lambda$ 5853, 6142, and 6497. These Ba II lines are also strong in SN 2005cs and SN 2009N, while they are weak in SN 1999em and SN 2012A.  In SN 2008in, the strength of \ion{Ba}{2} lines is medium among the comparison sample. \citet{Tur97D} suggested that these strong \ion{Ba}{2} lines are not owing to the over-abundance of Ba but it is probably a temperature effect based on the spectra modelling of faint SN IIP 1997D.  This might imply that a inverse relationship exist between the strength of \ion{Ba}{2} and the photospheric temperature of SNe IIP \citep{Hatano99,2009N}. The three components of \ion{Ca}{2} IR triplet can be also resolved (see also Figure \ref{<TriL>}) in SNe 2013am, 2005cs, and 2009N at this phase. In principle, formation of narrower spectral lines can relate with the slow moving ejecta and/or a narrower line-forming region.

\subsection{Ejecta Velocities}
\label{Ejecta}
In this subsection we examine the ejecta velocity of SN 2013am through H and some metal lines. The location of the blueshifted absorption minimum was
measured using both the Gaussian fit routine and the direct measurement of the absorption minimum, and the results were averaged. The upper panel
of Figure \ref{<VHa>} shows the velocity inferred from H$_{\alpha}$. As it can be seen, SN 2013am has a much lower ejecta velocity relative to the normal SNe IIP. At t$\sim$ +50 days after explosion, the expansion velocity of SN 2013am is $\sim$2000 km s$^{-1}$, while the corresponding value for the normal SNe IIP like SN 1999em is 5000-6000 km s$^{-1}$. A similarly low expansion velocity is also seen in SN 2005cs, with a velocity of $\sim$2300 km s$^{-1}$ at t$\approx$+44 days. Note that the expansion velocity of SN 2009N is similar to that of SN 1999em at $t < +50$days but it decreases rapidly after that and becomes comparable to SN 2005cs at t $\gtrsim$ 100 days. The spectral features and the velocity evolution of SN 2013am are overall similar to those of SN 2005cs.

We further examined the the ejecta velocity inferred from other lines of SN 2013am, as shown in the lower panel of Figure \ref{<VHa>}. We notice that this velocity relates to the strength of the corresponding lines owing to that the stronger lines typically form at a larger radius \citep{Mazz92}. In the earlier phase, H$\alpha$ is the strongest line and has the highest velocity; while at later phase, \ion{Ca}{2} IR triplet is the strongest feature and at the highest velocity.

\section{Discussion}
\label{sect:Discu}

\subsection{Absolute Magnitudes}
\label{subsect:AbsMag}

Applying the Tully-Fisher distance ($m-M$ = 30.54$\pm$0.40 mag; \citep{mM11}) and the reddening derived in Section 3.4 to the observed B$-$ and V$-$band peak magnitudes, we obtain the absolute magnitudes at maximum light as M$_B = -15.97= \pm 0.88$ mag and $M_V = -15.95\pm0.71$ mag when $R_V$=3.1 is adopted in the extinction corrections. On the other hand, a lower extinction ratio, i.e., R$_{V}$ = 1.4-1.5, has been suggested for some SNe Ia (i.e., \citealp{Wang09}) and even for SNe IIP (e.g., \citealp{Ext09,Oli10}) in light of the statistical study. Assuming R$_{V}$ = 1.4 for the extinction correction, the absolute magnitudes of SN 2013am drop to $-$15.07$\pm$0.61 mag in $B$ and $-15.05\pm0.48$ mag in $V$, which are close to those of SN 2005cs but are much lower than the normal SNe IIP.

\citet{Sanders} found a relation between the absolute peak brightness ($M_{peak}$) and plateau phase decline rate ($\beta_2$, defined in the fig.4 of \citealp{Sanders}) of SNe IIP, where the fainter SN IIP shows slower decline rate at the plateau phase. This relation is best fitted in the $r$-band via:
\begin{equation}
log_e\beta_2[r]=(-12.5\pm1.1)+M_{peak}[r](0.44\pm0.06).
\end{equation}
The decline rate of SN 2013am is $log_e\beta_2[r] = -5.84\pm0.20$  derived from the measurements listed in Table \ref{Tab:gri}, which yields M$(r)_{peak}=-15.14\pm0.46$ mag conforming to the result of $R_V$=1.4 (i.e.,  M$(r)_{peak} = -15.11\pm0.43$ mag). 

\subsection{Quasi-Bolometric light curve and synthesized $^{56}$Ni}
\label{subsect:bolo}
\begin{figure}[!th]
\centering
\includegraphics[width=8cm,angle=0]{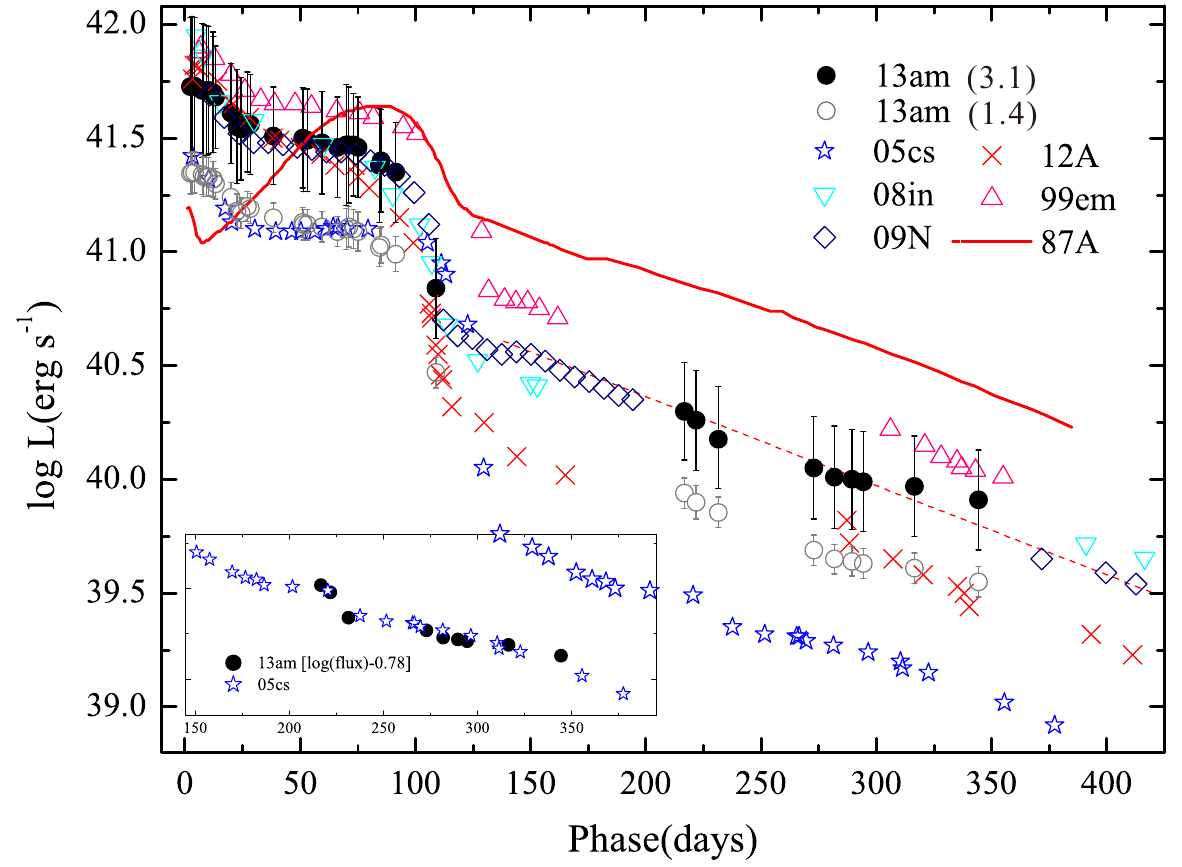}
 \caption{Comparison of the quasi-bolometric ($UBVRI$) light curve of SN 2013am with SNe 2005cs, 2008in, 2009N, 2012A, 1999em, and 1987A. The dashed line represents the slope of the Co$\rightarrow$Fe decay. The error bars shown for SN 2013am include the uncertainties in extinction correction, distance modulus, and photometry.The insets shows the comparison of scaled SN 2013am and SN 2005cs at late time.}
 \label{<bolo>}
\end{figure}

Using the UBVRI data presented in the previous sections, we computed the bolometric light curve of SN 2005cs. Figure \ref{<bolo>} shows the pseudo-bolometric ($UBVRI$) light curve of SN 2013am, together with those of SNe 1999em, 2005cs, 2008in, 2009N, 2012A, and 1987A.  As the late-time photometry was only available in $VRI$ bands for SN 2013am, we interpolate the missing $U$- and $B$-band flux from the well-observed SNe IIP such as SN 1999em and SN 1987A at some phases whenever it is necessary. This assumption is reasonable since different SNe IIP tend to show similar color evolution (except in $V - I$) at late times as shown in Figure \ref{<CC>}.

From the bolometric light curve, we find that SN 2013am reached a peak luminosity of $L=5.53^{+5.07}_{-2.54} \times 10^{41}$ erg s$^{-1}$) (for $R_V$ = 3.1) at about 4 days after explosion, which is comparable to SN 2008in and SN 2009N. While correcting for the extinction with $R_V$=1.4 yields a lower value $L=2.31^{+2.02}_{-1.09} \times 10^{41}$ erg s$^{-1}$, similar to that of SN 2005cs. Note that the bolometric light curve of SN 2013am declines more slowly relative to other comparison SNe in the first 2-3 weeks after explosion, which might be related to the slower decline in the $UV$ bands. On the other hand, the late-time evolution seems to be very similar to that of SN 2005cs, specially at period from $\sim$200 to $\sim$350 days after explosion (see inset of Figure \ref{<bolo>}). The decline rate on the exponential tail is estimated as 0.72$\pm0.08$ mag (100d)$^{-1}$, similar to that of SN 2005cs (i.e., 0.75$\pm0.05$ mag (100d)$^{-1}$), both are smaller than that expected from $^{56}$Co$\rightarrow$Fe decay.

The tail luminosity of the light curve can be used to estimate the mass of $^{56}$Ni ejected by the SN. Comparing the tail luminosity of SN 2013am with that of SN 1987A via the relation follows as \citep{Dan88,Woosley}:
\begin{equation}
M_{SN}(Ni)= 0.075\times \frac{L_{SN}}{L_{87A}} M_{\sun}
\end{equation}
From the above relation, we estimate that SN 2013am ejected $0.017^{+0.010}_{-0.007} $M$_{\sun}$ of $^{56}$Ni if $R_V$ = 3.1 is used, which is larger than typical value for low-velocity SNe IIP (i.e., $M(^{56}$Ni) $< 0.01$M$_{\sun}$, \citealp{Past04,Spiro14}) but is comparable to that ejected by SN 2008in (i.e., $M(^{56}$Ni) = 0.015 M$_{\sun}$, \citealp{Roy11}) and SN 2009N (i.e., $M(^{56}$Ni) = 0.02 M$_{\sun}$, \citealp{2009N}).

\citet{Ham03} presented an alternative way to calculate the tail luminosity using only the $V$-band light curve. This estimation is free from the uncertainty caused by interpolating the missing flux in the $U$- and $B$- bands for SN 2013am, which follows as
\begin{equation}
log_{10}L_t = \frac{-(V_t-A_V+BC)+5log_{10}D-8.14}{2.5}
\end{equation}
where $L_t$ is the tail luminosity in erg s$^{-1}$, $V_t$ is the $V$-band magnitude at the tail phase, $A_V$ is the total extinction in the $V$ band, and $D$
is the distance in cm. The bolometric correction is estimated as BC = 0.26$\pm$0.06 \citep{Ham01}. Once the tail luminosity is determined, the $^{56}$Ni mass can be estimated via
\begin{equation}
M_{Ni} = (7.866\times10^{-44}) L_t~exp\left[ \frac{t/(1+z)-6.1}{111.26} \right]M_{\sun}
\end{equation}
where $t$ is the phase after explosion. Inserting the $V$-band tail magnitude, the distance, and the extinction of SN 2013am into the above two formula, we obtain an estimate of the $^{56}$Ni mass as $0.015^{+0.011}_{-0.006}$M$_{\sun}$, which is in line with the above estimate.

\subsection{Diversities of SNe IIP}
\label{subsect:Diver}
\begin{figure}[!th]
\centering
\includegraphics[width=8cm,angle=0]{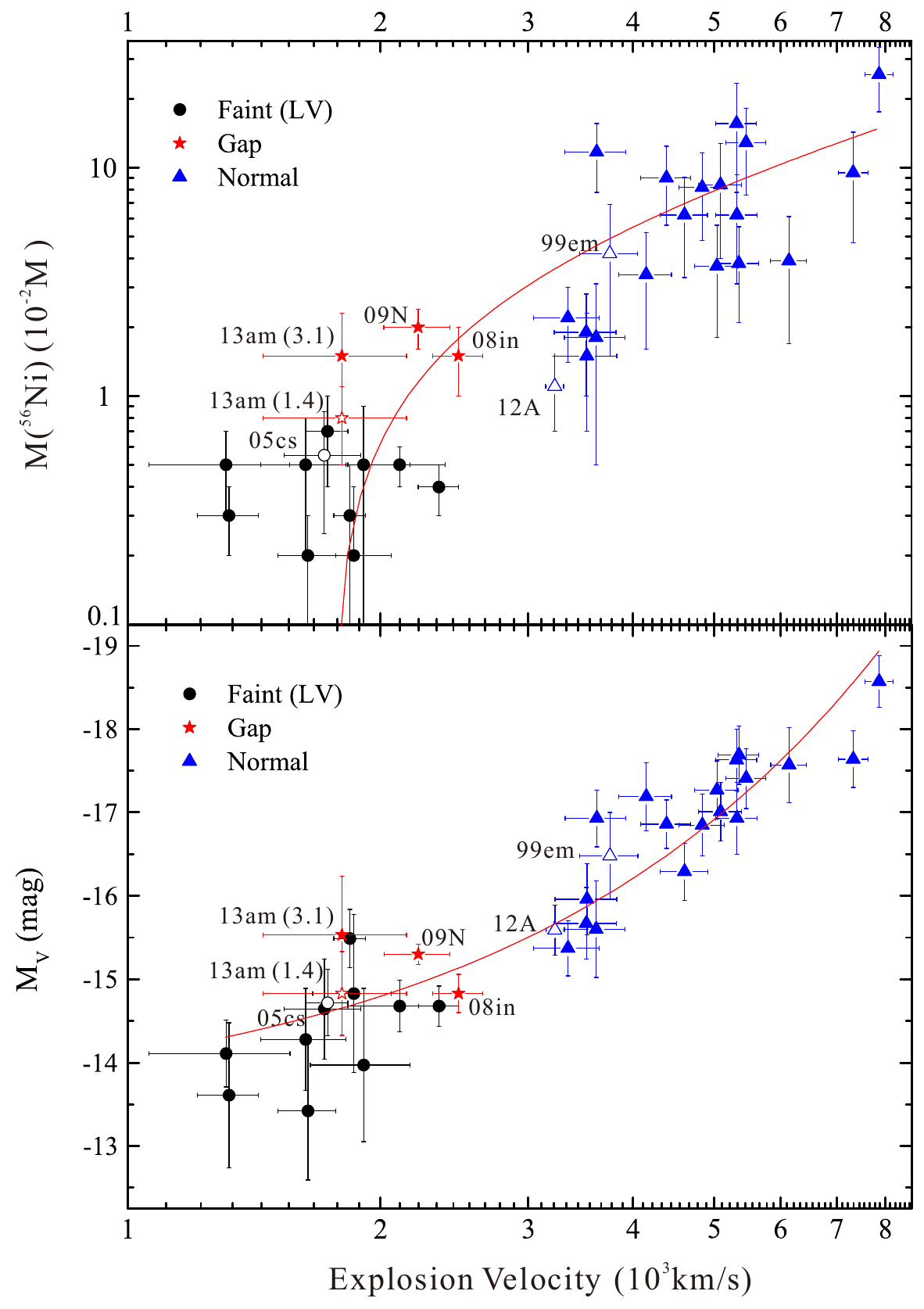}
 \caption{Comparison of various photometric and spectroscopic indicators from SN 2013am with those from other SNe IIP as measured by \citet{Ham03,Roy11,2009N,SN12A,Spiro14}. Upper: the velocity of \ion{Sc}{2} $\lambda$6246 or \ion{Fe}{2} $\lambda$5169 measured at $t\approx +50$ days after explosion (i.e., $v^{50}_{exp}$) vs. M($^{56}$Ni). Lower: $v^{50}_{exp}$ vs. the $V$-band absolute magnitudes measured at $t\approx +50$ days after explosion (i.e., $M_V^{50}$). The filled star represents SN 2013am dereddened with $R_V$=3.1, while the open star represents the case with $R_{V}$=1.4. SNe 2005cs, 2012A, and 1999em were plotted by opened signs for better display. The red lines represent the linear fit to the observed data.}
  \label{<ScNi>}
\end{figure}

To get better understanding of the properties of SN 2013am, we further examined its location in the correlations among $^{56}$Ni mass, expansion velocity of the ejecta and $V$-band absolute magnitude in the middle of the plateau (M$^{50}_{V}$).  Figure\ref{<ScNi>} shows such correlations for thirty four SNe IIP, including eleven low-luminosity or low-velocity objects from \citet{Spiro14}; twenty normal SNe IIP from \citet{Ham03}, and two transition events SN 2008in and SN 2009N. SN 1999em and SN 2012A are also included in the plot. Note that the velocity of normal SNe IIP were derived from \ion{Fe}{2} $\lambda$5169 \citep{Ham03} and the rest samples derived from \ion{Sc}{2} $\lambda$6246, both could indicate the photospheric velocity of SNe IIP.  Moreover, \citet{Maguire} suggested that \ion{Sc}{2} $\lambda$6246 is a better indicator of the photospheric velocity than \ion{Fe}{2} $\lambda$5169 owing to the  lower optical depth. On the other hand, \citet{Maguire} also indicated that the velocity of \ion{Sc}{2} $\lambda$6246 is very close to that of \ion{Fe}{2} $\lambda$5169 (i.e., $V_{Sc}$ = 0.95 $V_{Fe}$). Therefore, the original measures of  $V_{Fe}$ from \citet{Ham03} are adopted at here without correction.

As shown in Figure 11, the ejecta velocity of SNe IIP is largely scattered, ranging from $\sim$1500 km s$^{-1}$ to $\sim$8000 km s$^{-1}$. Significant scatter also exists in the plateau luminosity and the mass of $^{56}$Ni produced in the explosion. The wide range in luminosities and expansion velocities is clear manifestation of the great diversity of SNe IIP. Nevertheless, one can see that SNe IIP with larger ejecta velocities generally produce more $^{56}$Ni and are brighter during the plateau phase, which is the basis for the use of SNe IIP as distance indicators (e.g., \citealp{Ham03,Nug06,Ext09}). In the $^{56}$Ni--velocity space, SN 2013am is clearly separated from the low-velocity group and produced a nickel mass that is comparable to those of the ``gap-filler" objects like SN 2008in and SN 2009N. The presence of these transition objects confirms the hypothesis of a continuous distribution in explosion parameters of SNe IIP. In the M$^{50}_{V}$--velocity space, however, the separation of SN 2013am from other low-velocity SNe IIP is not distinct, though it appears still brighter for R$_{V}$ = 3.1. Moreover, we notice that the low-velocity sample of SNe IIP seems to show large scatter in synthesized nickel mass and plateau luminosity in comparison with those with higher ejecta velocities. More sample are needed to confirm wether the photometric diversity of SNe IIP  increases at low ejecta velocities.

On the other hand, the variations in the expansion velocities can be in principle related to the composition of the progenitor star, the envelope mass, and even the viewing angles to observe the explosions. At lower metallicity, the line-forming region will go deeper into the photosphere to have the same line opacity, and thus it results in smaller line velocities \citep{Lentz00}. Therefore, it is possible that the progenitor of SN 2013am has lower metallicity compared to SN 2008in or SN 2009in considering that they have similar tail luminosities. Moreover, the large envelope mass ejected in the explosion can also lead to smaller line velocity for the SNe IIP with the same explosion energy. On the other hand, the smaller ejecta velocity seen in SN 2013am might be also due to that it was viewed at an angle away from the polar direction in the explosion. This can be examined with the polarization observations made during the nebular phase.

\section{Summary}
\label{sect:con}
We present the optical and ultraviolet observations of the low-velocity SN IIP 2013am in nearby galaxy M65. The spectra of SN 2013am are characterized by narrow P-Cygni lines at lower velocities, and the overall evolution is very similar to that seen in the proto-type low-velocity II-P supernova 2005cs. However, SN 2013am has shorter ``plateau" duration and brighter tail compared to SN 2005cs, and it shows close resemblance to the transition objects like SN 2009N and SN 2008in.

Both photometric and spectroscopic methods suggest that SN 2013am suffered a significant reddening in the host galaxy (i.e., $E(B-V)\sim0.5\pm0.2$ mag). After corrections for the extinction using the above reddening and $R_{V}$=3.1, we find that SN 2013am has an absolute $V$-band magnitude of $-15.95\pm0.71$ mag in the middle of the plateau, a $UBVRI$ bolometric luminosity of $L=5.53^{+5.07}_{-2.54} \times 10^{41}$ erg s$^{-1}$) around the maximum light, and an ejected $^{56}$Ni mass of $0.016^{+0.010}_{-0.006}$M$_{\sun}$. These values are comparable to those derived for the ``gap-filler" events SN 2009N and SN 2008in, although the variation in R$_{V}$ may cause additional uncertainties in these estimations. The higher luminosity of SN 2013am, compared with SN 2005cs, is also consistent with the report that SNe IIP with shorter ``plateau" durations tend to be more luminous \citep{and14}. These results indicate that the velocity-luminosity relation may not hold for all SNe IIP.  The ejecta velocity may be related to variations in metallicity of the progenitor star, the envelope mass, and even the viewing angles to observe the explosions. A better understanding of the properties of the progenitors and polarization observations would help make further distinguish between the above factors.

\acknowledgments
We thank very much the anonymous referee for his/her constructive suggestions which helped to improve the paper a lot. We also acknowledge the support of the staff of the Li-Jiang 2.4-m telescope (LJT), Liverpool 2-m telescope (LT), Tsinghua-NAOC 80-cm telescope (TNT) and 3.58-m Telescopio Nazionale Galileo (TNG). Funding for the LJT has been provided by Chinese academe of science (CAS) and the People's Government of Yunnan Province. The LT is operated on the island of La Palma by Liverpool John Moores University in the Spanish Observatorio del Roque de los Muchachos of the Instituto de Astrofisica de Canarias with financial support from the UK Science and Technology Facilities Council. The TNT is owned by Tsinghua University and jointly operated by Tsinghua University and the National Astronomical Observatory of the Chinese Academy of Sciences (NAOC).  The TNG is operated on the island of La Palma by the Fundaci\'{o}n Galileo Galilei of the INAF (Istituto Nazionale di Astrofisica). The data of UVOT comes from the $Swift$ Data Center. We also very much thank A. Pastorello, R. Roy, K. Tak\'{a}ts, and L. Tomasella for providing their spectra of SN 2005cs, SN 2008in, SN 2009N, and SN 2012A.

J.J. Zhang is supported by the National Natural Science Foundation of China (NSFC, grant 11403096). The work of X.F. Wang is supported by the Major State Basic Research Development Program (2013CB834903), the NSFC (grants 11073013, 11178003, 11325313), Tsinghua University Initiative Scientific Research Program, and the Strategic Priority Research Program ``The Emergence of Cosmological Structures" of the Chinese Academy of Sciences (grant No. XDB09000000). The work of J. M. Bai is supported by the NSFC (grants 11133006, 11361140347) and the Strategic Priority Research Program ``The Emergence of Cosmological Structures" of the Chinese Academy of Sciences (grant No. XDB09000000). T.M. Zhang is supported by NSFC (grant 11203034).  J.G. Wang is supported by NSFC (grant 11303085), and the Western  Light Youth Project and the open Research Program of the key Laboratory for Research in Galaxies and Cosmology, CAS. X.L. Zhang  is supported by NSFC (grant 11203070). X.L. Wang is  supported by NSFC (grant 11103078).

 \clearpage


\begin{thebibliography}{}
\bibitem[Anderson et al. (2014)]{and14} Anderson, J., et al. 2014, \apj, 786, 67
\bibitem[Barbon et al.(1990)]{Bar25}Barbon R., Benetti S., Rosino L., Cappellaro E., Turatto M., 1990, \aap, 237, 79
\bibitem[Benetti et al.(2001)]{Be97D}Benetti, S., Turatto, M., Balberg, S., et al. 2001, \mnras, 322, 361
\bibitem[Benetti et al.(2013)]{Be13id} Benetti, S., Tomasella, A., Pastorello, E., et al. 2013, CBET 3440, 2
\bibitem[Brown et al.(2007)]{Browncs}Brown, P.J., Dessart, L., Holland, S.T., et al. 2007, \apj, 659, 1488
\bibitem[Brown et al.(2014)]{SOUSA}Brown, P.J.,  Breeveld, A. A., Holland, S, et al. 2014, AP\&SS, 301
\bibitem[Cardelli et al.(1989)]{Cardelli}Cardelli et al.  1989, ApJ, 345, 245
\bibitem[Catchpole et al.(1987)]{87A2}Catchpole, R.M., et al. 1987, \mnras, 229, 15
\bibitem[Catchpole et al.(1988)]{87A3}Catchpole, R.M., et al. 1988, \mnras, 231, 75
\bibitem[Cousins (1981)]{Cousins}Cousins, A. 1981, South African Astron. Obs. Circ., 6, 4
\bibitem[Danziger (1988)]{Dan88}Danziger I. J., 1988, snoy.conf, 3
\bibitem[Dessart et al.(2010)]{Dessart10}Dessart, L., Livne, E., \& Waldman, R. 2010, MNRAS, 408, 827
\bibitem[Elmhamdi et al.(2003)]{Elm_99em}Elmhamdi, A., Danziger, I. J.,  Chugai, N., 2003, \mnras, 338, 939
\bibitem[Fan et al.(2014)]{FYF} Fan, Y.F., Bai, J.M., Zhang, J.J., et al. 2014, Research in Astron. Astrophys, in press
\bibitem[Faran et al.(2014)]{Faran14}Faran, T., Poznanski, D., Filippenko, A.V., et al. 2014, arXiv:1404.0378
\bibitem[Fukugita et al.(1996)]{Fukugita96}Fukugita, M., Ichikawa, T., Gunn, J. E., et al. 1996, \aj, 111, 1748
\bibitem[Gehrels et al.(2004)]{Swift04}Gehrels, N., Chincarini, G., Giommi, P. et al. 2004, \apj, 611, 1005
\bibitem[Hamuy et al.(2001)]{Ham_99em}Hamuy, M., Pinto, P.A., Maza, J., et al. 2001, \apj, 558, 615
\bibitem[Hamuy (2001)]{Ham01} Hamuy, M. 2001, Ph.D. thesis, Univ. Arizona
\bibitem[Hamuy (2003)]{Ham03}Hamuy, M. 2003, \apj, 582, 905
\bibitem[Hatano et al.(1999)]{Hatano99}Hatano K., Branch D., Fisher A., Millard J., Baron E., 1999, ApJS, 121, 233
\bibitem[Houck \& Fransson (1996)]{HouckFransson}Houck, J., \& Fransson, C. 1996, \apj, 456, 811
\bibitem[Hillier \& Miller (1998)]{Hill98}Hillier D.J., \& Miller, D.L. 1998, \apj, 496, 407
\bibitem[Huang et al.(2012)]{TNT}Huang, F., Li, J.Z., Wang, X.F., et al. 2012, Research in Astron. Astrophys, 11, 1585
\bibitem[Jester et al.(2005)]{ugrizT}Jester, S.,  Schneider, D.P., Richards, G.T., et al. 2005, \aj, 130, 873
\bibitem[Jordi et al.(2006)]{Jordi06}Jordi, K., et al. 2006, \aap, 460, 339
\bibitem[Johnson et al.(1966)]{Johnson}Johnson, H., Iriarte, B., Mitchell, R., $\&$ Wisniewskj, W. 1966, Commun. Lunar Planet. Lab., 4, 99
\bibitem[Krisciunas et al.(2009)]{Kri09}Krisciunas, K., Hamuy, M., Suntzeff, N.B., et al. 2009, \apj, 137, 34
\bibitem[Landolt (1992)]{Landolt}Landolt, A. V. 1992, AJ, 104, 340
\bibitem[Lentz et al.(2000)]{Lentz00}Lentz, E.J., Baron, E., Branch, D., et al. 2000, \apj, 530, 966L
\bibitem[Leonard et al.(2001)]{Leo_99em} Leonard, D.C.,  Filippenko, A.V.,  Gates, E.L., et al. 2002, PASP, 114, 35
\bibitem[Maguire et al.(2010)]{Maguire}Maguire, K., Di Carlo, E., Smartt, S.J., et al. 2010, \mnras, 404, 981
\bibitem[Mazzali et al.(1992)]{Mazz92}Mazzali, P.A., Lucy, L.B., $\&$ Butler, K., 1992, \aap, 258, 399
\bibitem[Menzies et al.(1987)]{87A1}Menzies, J.W., et al. 1987, \mnras, 227, 39
\bibitem[Nasonova et al. (2011)]{mM11}Nasonova, O. G., de Freitas Pacheco, J. A., $\&$ Karachentsev, I. D. 2011, AA, 532, 104
\bibitem[Nugent et al.(2006)]{Nug06}Nugent, P., Sullivan, M.,  Ellis, R., 2006, \apj, 645, 841
\bibitem[Olivares et al.(2010)]{Oli10}Olivares, F., Hamuy, M., Pignata, G., et al. 2010, \apj, 715, 833
\bibitem[Pastorello et al.(2004)]{Past04}Pastorello,A., Zampieri, L., Turatto, M., et al. 2004, \mnras, 347,74
\bibitem[Pastorello et al.(2006)]{05csa}Pastorello,A., Sauer, D., Taubenberger, S., et al. 2006, \mnras, 370, 1752
\bibitem[Pastorello et al.(2009)]{05csb}Pastorello, A., Valenti, S.,  Zampieri, L., et al. 2009, \mnras, 394, 2266
\bibitem[Poole et al.(2008)]{UVOTcali}Poole, T., Breeveld,  A., Page, M., et al. 2008, \mnras, 383, 627
\bibitem[Poznanski et al.(2009)]{Ext09}Poznanski, D., et al. 2009, \apj, 694, 1067
\bibitem[Poznanski et al.(2011)]{Ext11}Poznanski, D., Ganeshalingam, M., Silverman, J.M., $\&$ Filippenko, A.V. 2011, \mnras, 415, L81
\bibitem[Pritchard et al.(2014)]{Prit}Pritchard, T.A., Roming, P., Brown, P., et al. 2014, \apj, 787, 157
\bibitem[Roming et al.(2005)]{UVOT05} Roming, P., Kennedy, T., Mason, K., et al. 2005, Space Sci. Rev., 120, 95
\bibitem[Roy et al.(2011)]{Roy11}Roy R., et al., 2011, \apj, 736, 76
\bibitem[Sanders et al.(2014)]{Sanders}Sanders, N.E., Soderberg, A.M.,  Gezari, S., et al. 2014, arXiv:1404.2004
\bibitem[Schlegel et al.(1998)]{Schle98}Schlegel, D., Finkbeiner, D., and Davis, M., 1998, \apj, 500, 525
\bibitem[Smartt (2009)]{CC09}Smartt, S.J. 2009, \araa. 47, 63
\bibitem[Spiro et al.(2014)]{Spiro14}Spiro, S., Pastorello, A., Pumo, M.L., et al. 2014, \mnras, 439, 2873
\bibitem[Stetson(1987)]{Stetson}Stetson, P. 1987, \pasp, 99, 191
\bibitem[Sugano(2013)]{Suga13} Sugano, M., 2013, CEBT  3440, 1
\bibitem[Tak\'{a}ts et al.(2014)]{2009N}Tak\'{a}ts, K., Pumo, M. L., Elias-Rosa, N., et al. 2014, \mnras, 438, 368
\bibitem[Tomasella et al.(2013)]{SN12A}Tomasella, L., Cappellaro, E., Fraser, M., et al. 2013, \mnras, 434, 1636
\bibitem[Turatto et al.(1998)]{Tur97D}Turatto, M., Mazzali, P.A., Young, T.R., et al. 1998, \apj, 498, L129
\bibitem[Turatto et al.(2003)]{Tura16} Turatto, M., Benetti, S.,  \& Cappellaro, E., 2003, in From Twilight to Highlight: The Physics of Supernovae, Hillebrandt, W.,  Leibundgut, B., eds., p.200
\bibitem[Utrobin (2007)]{Utrobin}Utrobin, V. P. 2007, \aap, 461, 233
\bibitem[Wang et al.(2008)]{Wang08} Wang, X., et al. 2008, \apj, 675, 626
\bibitem[Wang et al. (2009)]{Wang09} Wang, X., et al. 2009, \apj, 699, L139
\bibitem[Woosley et al(1989)]{Woosley}Woosley S. E., Hartmann D., Pinto P. A.1989, ApJ, 346, 395
\bibitem[Whitelock et al.(1988)]{87A4}Whitelock, P.A., et al. 1988, \mnras, 234, 5
\bibitem[Yaron et al.(2013)]{ATel4910}Yaron, O., Gal-Yam, A., Fox, O.D., et al. 2013, ATel 4910.
\bibitem[Zhang et al.(2012)]{zhang12} Zhang, J.J., Fan. Y.F., Chang, L., et al., 2012, Astronomical Research Technology, 9,411
\bibitem[Zhang et al.(2014)]{JJzhang14} Zhang, J.J, Wang, X.F., Bai, J.M., et al., 2014, \aj, 148, 1

\end{thebibliography}
\end{document}